\documentclass[12pt]{article}
\usepackage{epsfig,amssymb,amsmath,psfrag}

\def\eqn#1{eq.~(\ref{#1})}

\def\eqns#1#2{eqs.~(\ref{#1}) and~(\ref{#2})}

\def\cN{{\mathcal N}}

\def\polylog{ {\rm Li} }
\def\ln{ \log }
\def\ws{ w^{*} }
\def\ty{\tilde{y}}

\def\be{\begin{equation}}
\def\ee{\end{equation}}
\def\bea{\begin{eqnarray}}
\def\eea{\end{eqnarray}}

\begin{document}
\thispagestyle{empty}

\begin{center}
CERN--PH--TH/2011/189\hskip1cm
SLAC--PUB--14528 \hskip1cm
LAPTH-029/11\\
HU-EP-11-38 \hskip1cm NSF-KITP-11-176\hskip1cm
\end{center}

\begingroup\centering
{\Large\bfseries\mathversion{bold}
Bootstrapping the three-loop hexagon \par}%
\vspace{7mm}

\begingroup\scshape\large
Lance~J.~Dixon$^{(1,2)}$, James~M.~Drummond$^{(1,3)}$\\
and Johannes M.~Henn$^{(4,5)}$\\
\endgroup
\vspace{8mm}
\begingroup\small
$^{(1)}$ \emph{PH-TH Division, CERN, Geneva, Switzerland} \\
$^{(2)}$ \emph{SLAC National Accelerator Laboratory,
Stanford University, Stanford, CA 94309, USA} \\
$^{(3)}$ \emph{LAPTH, Universit\'e de Savoie, CNRS,
B.P. 110, F-74941 Annecy-le-Vieux Cedex, France}\\
$^{(4)}$ \emph{Institut f\"ur Physik, Humboldt-Universit\"at
zu Berlin, Newtonstra{\ss}e 15, D-12489 Berlin, Germany}\\
$^{(5)}$ \emph{Kavli Institute for Theoretical Physics,
University of California, Santa Barbara, CA 93106, USA}
\endgroup

\vspace{3cm}

\textbf{Abstract}\vspace{5mm}\par
\begin{minipage}{14.7cm}
We consider the hexagonal Wilson loop dual to the six-point MHV
amplitude in planar $\mathcal{N}=4$ super Yang-Mills theory.
We apply constraints from the operator product expansion in the
near-collinear limit to the symbol of the remainder function at three loops.
Using these constraints, and assuming a natural ansatz for the symbol's
entries, we determine the symbol up to just two undetermined constants.
In the multi-Regge limit, both constants drop out from the
symbol, enabling us to make a non-trivial confirmation of the BFKL
prediction for the leading-log approximation. This result
provides a strong consistency check of both our ansatz for the symbol
and the duality between Wilson loops and MHV amplitudes.
Furthermore, we predict the form of the full three-loop
remainder function in the multi-Regge limit, beyond the leading-log
approximation, up to a few constants representing terms not detected by
the symbol.  Our results confirm an all-loop prediction
for the real part of the remainder function in multi-Regge
$3\to3$ scattering.
In the multi-Regge limit, our result for the remainder
function can be expressed entirely in terms of classical polylogarithms.
For generic six-point kinematics other functions are required.
\end{minipage}\par
\endgroup

\newpage


\section{Introduction and outline}

Scattering amplitudes in $\mathcal{N}=4$ super Yang-Mills theory (SYM)
have fascinating properties, especially in the planar limit. One of
their most surprising properties is an equivalence with
certain light-like Wilson loop configurations, for which there
is strong empirical evidence at weak coupling, as well as general
arguments originating from strong coupling~\cite{Alday:2007hr,%
Drummond:2007aua,Brandhuber:2007yx,Drummond:2007cf,Drummond:2007bm,%
Bern:2008ap,Drummond:2008aq,Berkovits:2008ic}. The equivalence
relates the suitably-defined finite parts of maximally-helicity-violating
(MHV) scattering amplitudes to
the finite parts of Wilson loops evaluated on a null polygonal contour
in dual (or region) space.  The edges of the polygon are defined by the
gluon momenta $p_i^\mu$ via
\be
p_i^\mu = x_i^\mu - x_{i+1}^\mu\,.
\ee
The contour has
corners (or cusps) at the points $x_i$.  The equivalence between
amplitudes and Wilson loops implies that the analytic
properties of Wilson loops in the dual space can be identified with
those of scattering amplitudes in momentum space.

Wilson loops in a conformal field theory exhibit conformal
symmetry. The null polygonal Wilson loops related to scattering
amplitudes are ultraviolet divergent due to the presence of cusps on
the contour. Nonetheless they still obey a conformal Ward
identity~\cite{Drummond:2007cf,Drummond:2007au}.  This identity
can be simply stated as follows.  We write the logarithm of the Wilson
loop with $n$ cusps as a sum of divergent and finite terms,
\be \log W_n = [{\rm UV\ divergent}]_n +
F_n^{\rm WL}\,.\ee
The Ward identity for the finite part is then
\be
K^\mu F_n^{\rm WL} = - \frac{\gamma_K}{2} \,
\sum_{i=1}^{n} (2x^\mu_i - x^\mu_{i-1} - x^\mu_{i+1})
\, \log x_{i-1,i+1}^2\,,
\label{confWI}
\ee
where $x_{i,j}=x_{i}-x_{j}$ and $x_{i+n} \equiv x_{i}$, and
$K^\mu$ are the generators of (dual) special conformal transformations,
\be \label{Kdef}
K^\mu = \sum_{i=1}^n \biggl[ 2 x_i^\mu x_i^\nu 
\frac{\partial}{\partial x_i^\nu}
- x_i^2 \frac{\partial}{\partial x_{i\,\mu}} \biggr] \,.
\ee
The cusp anomalous dimension~\cite{Korchemskaya:1992je} $\gamma_K$
is predicted to all orders in the coupling constant~\cite{Beisert:2006ez}.

The Ward identity~(\ref{confWI}) fixes $F_n^{\rm WL}$, up to functions
of conformally invariant cross ratios.  Below six points there are no
such cross-ratios and the solution is unique up to an additive
constant.  In fact this solution coincides precisely with the BDS
ansatz~\cite{Bern:2005iz} for the finite part of MHV scattering
amplitudes.  At six points and beyond there are
cross ratios, so the solution is not unique.
The BDS ansatz still provides a particular solution to the
Ward identity for all $n$, but it does not give the complete answer.
A convenient way of writing the solution to the Ward identity is then
\be
F_n^{\rm WL} = \gamma_K \, F^{\rm WL}_{n\,,{\rm 1-loop}} + R_n\,.
\label{WardIdSoln}
\ee
Here $F^{\rm WL}_{n\,,{\rm 1-loop}}$ is the one-loop result for
$F_n^{\rm WL}$, while $R_n$ is the `remainder function', which is a
function only of conformal cross ratios and becomes non-vanishing at
two loops~\cite{Drummond:2007bm}.  In terms of the loop expansion
parameter $a \equiv g^2 N_c/(8\pi^2)$, the remainder function is
expanded as
\be
R_n\ =\ \sum_{l=2}^\infty a^l \, R_n^{(l)} \,.
\label{Rnnorm}
\ee 

At six points, the remainder
function depends on three dual conformal cross ratios,
\bea
u = \frac{x_{13}^2 x_{46}^2}{x_{14}^2 x_{36}^2}
= \frac{ s_{12} s_{45} }{ s_{123} s_{345} }\,, \qquad
v = \frac{x_{24}^2 x_{15}^2}{x_{25}^2 x_{14}^2}
= \frac{ s_{23} s_{56} }{ s_{234} s_{456} }\,, \qquad
w = \frac{x_{35}^2 x_{26}^2}{x_{36}^2 x_{25}^2}
= \frac{ s_{34} s_{61} }{ s_{345} s_{561} }\,,
\eea
which are in turn built from the Lorentz invariants
$s_{i,j} = (p_i + p_{j})^2$ and $s_{i,j,k}=(p_{i}+p_{j}+p_{k})^2$.
The gluon momenta for the scattering process, $p_{i}^{\mu}$ with
$i=1,2,\ldots,6$, satisfy the on-shell conditions $p_{i}^2=0$.

The conformal symmetry of the Wilson loop implies that the dual planar
MHV amplitudes exhibit `dual conformal symmetry'. This symmetry has
been observed in the form of the scattering amplitudes in many guises:
the form of the integrals in the perturbative
expansion~\cite{Drummond:2006rz,Bern:2006ew,FiveLoop};
the background isometry of the AdS sigma model after
T-duality~\cite{Alday:2007hr,Berkovits:2008ic,BRTWdualsuper};
the structure of tree-level amplitudes, where it extends to dual
superconformal symmetry~\cite{Drummond:2008vq}, 
and combines with the original Lagrangian superconformal symmetry to
form a Yangian symmetry~\cite{Drummond:2009fd}; 
the structure of the scattering amplitudes on
the Coulomb branch~\cite{Alday:2009zm} and in higher
dimensions~\cite{Alday:2009zm,Bern:2010qa,CaronHuot:2010rj,Dennen:2010dh};
and the form of the on-shell recursion relations for the four-dimensional
planar integrand~\cite{ArkaniHamed:2010kv}. Many review articles are
available on different aspects of all of these developments, including
refs.~\cite{Radcor07,Alday:2008yw,Henn:2009bd,Henn:2011xk,Adamo:2011pv,%
Drummond:2010ep,Drummond:2010km,Bargheer:2011mm}. 
For the purposes of this paper the important point is simply
that the Ward identity~(\ref{confWI}) requires the function $R_6$ to depend
only on the invariant cross ratios $u$, $v$ and $w$.

Much recent progress~\cite{DelDuca:2009au,DelDuca:2010zg,DelDuca:2010zp,%
Goncharov:2010jf,Heslop:2010kq}
has focused on understanding the structure of the remainder function,
in part due to the fact that this same function governs the structure
of scattering amplitudes, both at strong coupling~\cite{Alday:2007hr}
and in the MHV sector in perturbation
theory~\cite{Drummond:2007bm,Bern:2008ap,Drummond:2008aq}. Understanding its
form then promises to greatly enhance our understanding of scattering
amplitudes in general.  A very important result in this direction was
the analytic calculation of the Feynman integrals appearing at two loops
in the hexagonal Wilson loop~\cite{DelDuca:2009au,DelDuca:2010zg}, 
which provided a closed-form expression for the remainder function in
terms of (many) multi-dimensional polylogarithms, or Goncharov
polylogarithms.  Remarkably, this seemingly complicated expression could
be dramatically simplified into a few lines of classical
polylogarithms~\cite{Goncharov:2010jf}.  An important tool for finding 
such a compact form of the two-loop, six-point remainder function is the
notion of the symbol of a transcendental function~\cite{symbols}.  The
symbol is a quantity which preserves the underlying algebraic nature of the
function, while forgetting about certain analytic properties, such as the
particular branch cut on which the function should be evaluated.
Complicated identities between polylogarithms reduce to simple algebraic
relations at the level of symbols.  The symbol can therefore be a key
step in discerning the analytic structure of amplitudes.  For example,
a conjecture has been made recently for the symbol of the two-loop
remainder function for an arbitrary number of points~\cite{CaronHuot:2011ky}.
Of course, eventually one would like to reconstruct the actual function
represented by the symbol.

Another important property of polygonal Wilson loops is that they
should respect a particular operator product expansion (OPE) in the
region where several consecutive edges are nearly
collinear~\cite{Alday:2010ku,Gaiotto:2010fk}.
This idea has recently been used to
argue that at two loops the hexagon remainder function can be uniquely
fixed from the knowledge only of the leading corrections to the
energies of the exchanged states in the OPE~\cite{Gaiotto:2011dt}. The
OPE has also recently been used to address the same problem for Wilson
loops with more than six sides~\cite{Sever:2011pc}, and for super Wilson
loops associated with non-MHV amplitudes~\cite{Sever:2011da}.

An important kinematical limit of higher-point scattering amplitudes is
the multi-Regge limit.  This limit is a generalization of the high-energy
limit of four-point scattering, but is one in which multiple parameters
can survive, related to the ordering of the final-state particles in rapidity.
For the MHV amplitudes in planar $\mathcal{N}=4$ super Yang-Mills theory,
this structure has been explored in several
papers~\cite{Bartels:2008ce,Bartels:2008sc,Lipatov:2010qf,%
Lipatov:2010qg,Lipatov:2010ad,Bartels:2010tx,Bartels:2011xy}.
Indeed, this limit provided early evidence that the BDS ansatz needed to
be corrected at two loops, starting with the six-point
amplitude~\cite{Bartels:2008ce}.  While the remainder function $R_6(u,v,w)$
vanishes in the Euclidean version of multi-Regge
kinematics~\cite{Brower:2008nm,Brower:2008ia,DelDuca:2008jg},
in the physical region its discontinuity is nonzero and can be analysed.
When dual conformal invariance holds, this discontinuity depends
nontrivially on two dimensionless variables, rather than the three
variables $u$, $v$ and $w$ characterizing generic kinematics.

A consequence of the duality between MHV amplitudes and Wilson loops is
that the multi-Regge behaviour of the amplitude should be consistent
with the OPE behaviour of the Wilson loop in the near-collinear limit.
That is, there is a further limit one can take of the multi-Regge kinematics
which is collinear in nature. This combined limit was studied
recently~\cite{Bartels:2011xy}, and it was shown that constraints from
the two limits pass a self-consistency test.

In this paper, inspired by all these exciting recent developments,
we will make an ansatz for the symbol of the three-loop hexagon remainder
function, $R_6^{(3)}(u,v,w)$, which is heavily constrained by the structures
described above.  We are able to apply all of the physical requirements,
such as the correct collinear behaviour, OPE expansion, multi-Regge limits
and so on, at the level of the symbol.  The correct near-collinear behaviour,
governed by the OPE expansion, is one of the strongest constraints on our
ansatz.  It is quite non-trivial that there is a consistent solution to the 
combined constraints.   For general kinematics, the solution for the symbol
is not unique, but contains 26 arbitrary constants.  However, all but three
of these parameters are irrelevant in the multi-Regge limit. 
Analysing the symbol in this limit, and imposing consistency with the
leading logarithmic prediction~\cite{Lipatov:2010ad}, we find that 
two of the three parameters relevant in this limit can be fixed.
Hence the symbol is completely fixed, up to a single constant parameter,
in this regime.  An additional constraint enables us to 
show that this extra constant parameter actually vanishes. The latter
constraint is an all-loop-order prediction~\cite{Bartels:2010tx}
concerning the behaviour of the real part of the remainder function
in the multi-Regge limit, after analytic continuation to $3\to3$ kinematics.
We have found functions corresponding to the symbol
in this limit, and we constrain the possible beyond-the-symbol
ambiguities in term of a few additional constants.  These functions
are all expressible in terms of classical logarithms and polylogarithms.
Thus we are able to make new, rather non-trivial predictions for the
next-to-leading and next-to-next-to-leading logarithmic approximations
to the scattering of six gluons at three loops in the multi-Regge limit.

We then examine the implications of imposing a further requirement on
the form of the final entries of our symbol. This restriction constrains
the derivatives of the remainder function.  It can be motivated by
the differential equations obeyed by
one-loop~\cite{Drummond:2010cz,Dixon:2011ng,DelDuca:2011wh}
and multi-loop integrals~\cite{DDHtoappear} related to scattering
amplitudes in planar $\cN=4$ super-Yang-Mills theory.
The same restriction has also been identified within a supersymmetric
formulation of the Wilson loop~\cite{CaronHuot:2011ky}.
We find that imposing this final-entry condition fixes the symbol
completely up to just two free parameters, and furthermore it determines
the symbol uniquely in the multi-Regge limit, and consistently with
the all-loop-order prediction for $3\to3$ scattering.

The paper is organised as follows. Section~\ref{sect-symbols} contains a
brief review of pure functions and properties of their associated symbols.
In section~\ref{sect-constraints} we make an ansatz for the symbol of
the remainder function of a particular, natural form, and we describe
the constraints that it must satisfy in order to be consistent.
In section~\ref{sect-OPE-constraints} we discuss the most involved
constraints, namely the ones coming from certain leading terms in the
OPE expansion.  Our focus
is on the interesting case of the hexagon at three loops.  We find that our
ansatz is consistent with all of the constraints we apply, and indeed
there is a 26-parameter solution at this stage.
In section~\ref{sect-Regge-predictions} we analyse our symbol in
multi-Regge kinematics, and produce new expressions for the next-to-leading
and next-to-next-to-leading logarithmic approximations at three loops.
In section~\ref{finalentry} we discuss the condition we impose on
the final entry of the symbol, and describe how it reduces our
ansatz to just two free parameters.  We also remark that for generic
values of $u$, $v$ and $w$, the three-loop remainder function cannot
be described in terms of classical polylogarithms, in contrast to what
happened at two loops.  In section~\ref{sect-conclusions} we present
our conclusions and give a brief outlook.  Three appendices give some
useful relations between different sets of kinematic variables,
as well as an alternate representation of the logarithmic coefficients
in the multi-Regge limit.

In additional files accompanying this article, as both Mathematica
notebooks and plain text files, we provide
the symbol for the three-loop remainder function, after imposing 
the final-entry constraint.  We also provide the symbols associated with
the remainder function in the multi-Regge limit.


\section{Pure functions and symbols}
\label{sect-symbols}

The remainder function of $\cN=4$ SYM is expected to be described in
terms of {\sl pure} functions. We define a pure function of degree
(or weight) $k$ recursively, by demanding that its differential satisfies
\be
d\, f^{(k)} = \sum_{r} f_r^{(k-1)} d \log \phi_r\,.
\label{pure}
\ee
Here the sum over $r$ is finite and $\phi_r$ are algebraic functions.
This recursive definition is for all positive $k$; the only degree zero
pure functions are constants. The definition~(\ref{pure})
includes logarithms and classical polylogarithms, as well as other
iterated integrals, such as harmonic polylogarithms of
one~\cite{Remiddi:1999ew}
or more~\cite{Gehrmann:2000zt,Gehrmann:2001ck,Gehrmann:2001pz,Maitre:2005uu}
variables.

The {\sl symbol}~\cite{symbols} ${\mathcal S}(f)$ of a pure function
$f$ is defined recursively with respect to~\eqn{pure},
\be 
{\mathcal S}( f^{(k)} )
= \sum_r {\mathcal S}( f_r^{(k-1)} ) \otimes \phi_r\,.
\label{symbol}
\ee
If we continue this process until we reach degree 0, we find that 
${\mathcal S}( f^{(k)} )$ is an element of the
$k$-fold tensor product of the space of algebraic functions,
\be
{\mathcal S}( f^{(k)} ) =
\sum_{\vec \alpha} \phi_{\alpha_1} \otimes \ldots \otimes \phi_{\alpha_k}\,,
\ee
where $\vec \alpha \equiv \{ \alpha_1,\ldots,\alpha_k \}$.
The symbol of a function does not contain all the information about the
function.  In particular, it loses information about which logarithmic
branch the integrand of an iterated integral is on, at each stage of
integration.  It also does not detect functions that are
transcendental constants multiplied by pure functions of lower degree.
(That is, such functions have zero symbol.)  The symbol therefore
corresponds to an equivalence class of functions that differ in these
aspects.  Nevertheless, the symbol is extremely useful, because
complicated identities between transcendental functions defined by
iterated integrals become simple algebraic identities.

If a symbol can be expressed as a sum of terms, with all entries in each
term belonging to a given set of variables, then we say that the
symbol can be factorised in terms of that set of variables.
From the definition of the symbol, a term containing an entry which
is a product can be split into the sum of two terms, according to
\be
\ldots \otimes \phi_1 \phi_2 \otimes \ldots
\ \ =\ \ \ldots \otimes \phi_1 \otimes \ldots\ \
+\ \ \ldots \otimes \phi_2 \otimes \ldots \,.
\ee
Performing this factorisation is usually necessary to identify all
algebraic relations between terms.  It is often necessary to perform
the step again after taking a kinematic limit, because the algebraic
relations in the limit are different than for generic kinematics.

The elements of the symbol are not all independent. In particular the
integrability condition $d^2 f^{(k)}=0$ for any function implies
relations among the different elements.  These relations
can be described simply:  One picks two adjacent slots in the symbol
$\phi_{\alpha_i} \otimes \phi_{\alpha_{i+1}}$ and replaces the
corresponding elements by the wedge product
$d \log \phi_{\alpha_i} \wedge d \log \phi_{\alpha_{i+1}}$ in every
term. The resulting expression must vanish.

It is very helpful in our analysis to consider the discontinuities of
the functions involved.  The symbol makes clear the locations of
the discontinuities of the function. If we have
\be {\mathcal S}( f^{(k)} ) = \sum_{\vec \alpha} 
\phi_{\alpha_1} \otimes \ldots \otimes \phi_{\alpha_k} \,,
\ee
then the degree $k$ function $f^{(k)}$ will have a branch cut starting at
$\phi_{\alpha_1}=0$. The discontinuity across
this branch cut, denoted by $\Delta_{\phi_{\alpha_1}} f^{(k)}$,
will also be a pure function, of degree $k-1$.  Its symbol is found
by clipping the first element off the symbol for $f^{(k)}$:
\be
{\mathcal S}( \Delta_{\phi_{\alpha_1}} f^{(k)} ) = \sum_{\vec \alpha}
\phi_{\alpha_2} \otimes \ldots \otimes \phi_{\alpha_k}\,.
\label{discsymbol}
\ee
In general, taking discontinuities commutes with taking derivatives.


\section{Constraining the three-loop remainder function}
\label{sect-constraints}

We will now describe a procedure for constraining the form of the
remainder function based on a plausible ansatz for its symbol. Our
experience with six-point integrals in both four and six
dimensions~\cite{Drummond:2010cz,Dixon:2011ng,DelDuca:2011ne} is that
their symbols are always formed of terms with entries drawn from the
following set of nine elements,
\be
\{ u, v, w, 1-u, 1-v, 1-w, y_u, y_v, y_w \}\,.
\label{ansatz}
\ee
Here we use the notation
\be \label{variablesy}
       y_u = \frac{u - z_+}{u - z_-}\,, 
\qquad y_v = \frac{v - z_+}{v - z_-}\,,
\qquad y_w = \frac{w - z_+}{w - z_-}\,,
\ee
where 
\be
z_\pm = \frac{1}{2}\left[-1+u+v+w\pm\sqrt{\Delta}\right]\,,
\qquad \Delta = (1-u-v-w)^2 - 4 u v w\,.
\ee
Thus our ansatz for the remainder function at $l$ loops will be the
most general symbol of degree $2l$ that we can make from the above set
of nine elements.  That is, we assume that the symbol for the remainder
function can be factorised in terms of the set~(\ref{ansatz}).

We can also write the cross ratios in terms of ratios of
two-brackets of $\mathbb{CP}^1$ variables $w_i$,
\be
       u = \frac{(23)(56)}{(25)(36)}\,, 
\qquad v = \frac{(34)(61)}{(36)(41)}\,,
\qquad w = \frac{(45)(12)}{(41)(52)}\,,
\label{uvw_to_wi}
\ee
where $(ij)=-(ji)=\epsilon_{ab}w^{a}_{i} w^{b}_{j}$.  In these
variables, $\Delta$ is a perfect square,
\be
\sqrt\Delta = \pm \frac{(12)(34)(56)+(23)(45)(61)}{(14)(25)(36)} \,.
\ee
Taking the positive branch of the square root, and using the
Schouten identity for the two-brackets, we have
\bea
        1 - u &=& \frac{(35)(26)}{(25)(36)}\,, 
\qquad   1 - v = \frac{(46)(31)}{(36)(41)}\,,
\qquad   1 - w = \frac{(51)(42)}{(41)(52)}\,,
\label{omuvw_to_wi}\\
       y_u &=& \frac{(23)(46)(15)}{(56)(13)(24)}\,,
\qquad  y_v = \frac{(61)(24)(35)}{(34)(51)(26)}\,,
\qquad  y_w = \frac{(45)(62)(31)}{(12)(35)(46)}\,. 
\label{y_to_wi}
\eea
Note that under a cyclic permutation, $w_i \to w_{i+1}$, with
indices modulo 6, the sign of $\sqrt{\Delta}$ flips, 
$\sqrt{\Delta} \to -\sqrt{\Delta}$.  So the $y$ variables permute
as $y_u \to 1/y_v \to y_w \to 1/y_u$.  This inversion will not affect
the symmetry properties of the parity-even functions and symbols in which
we are interested, which involve even numbers of $y$ variables.

From eqs.~(\ref{uvw_to_wi}), (\ref{omuvw_to_wi}) and (\ref{y_to_wi})
we see that our ansatz is equivalent
to saying that the symbol can be factorised in terms of two-brackets $(ij)$
(or equivalently momentum-twistor four-brackets) at the
six-point level.  (There are 15 two-brackets $(ij)$, but only combinations
that are invariant under rescaling of individual $w_i$ coordinates
are allowed, which reduces the number of independent combinations to
the nine exhibited in 
eqs.~(\ref{uvw_to_wi}), (\ref{omuvw_to_wi}) and (\ref{y_to_wi}).)
Note that we can fix a coordinate choice $w_i=(1,z_i)$, where
these variables coincide with the $z_i$ variables of
ref.~\cite{Goncharov:2010jf}, so that our ansatz is also equivalent
to assuming that the symbol can be factorised in terms of
differences of the $z_i$.   The form of our ansatz is certainly
sufficient at the two-loop level, because the remainder function is
explicitly known~\cite{Bern:2008ap,Drummond:2008aq,%
DelDuca:2009au,DelDuca:2010zg,Goncharov:2010jf},
and its symbol is indeed of this form~\cite{Goncharov:2010jf}.
In the above variables, it is given by
\bea
&&\mathcal{S}(R_6^{(2)})\nonumber\\
&&\hskip0.0cm = - \frac{1}{8} \Bigl\{
\null  \Bigl[ u \otimes (1-u) \otimes \frac{u}{(1-u)^2}
    + 2 \, \bigl( u \otimes v + v \otimes u ) \otimes \frac{w}{1-v}
    + 2 \, v \otimes \frac{w}{1-v} \otimes u \Bigr] 
\otimes \frac{u}{1-u}
\nonumber\\
&&\hskip0.8cm\null
+ \Bigl[ u \otimes (1-u) \otimes y_u y_v y_w
       - 2 \, u \otimes v \otimes y_w \Bigr] \otimes y_u y_v y_w \Bigr\}
\ +\ \hbox{permutations} \,,
\label{R62symb}
\eea
where the sum is over the 6 permutations of $u$, $v$ and $w$, which
correspondingly permute $y_u$, $y_v$ and $y_w$.

What constraints should the symbol of the remainder function obey?
\begin{itemize}
\item It should be integrable, {\it i.e.}~it should actually be the symbol
of a function.
\item The first entry in any term of the symbol should be a
cross ratio $u$, $v$ or $w$.  The leading entries describe the
locations of the discontinuities of the function, which can only
originate at $x_{ij}^2=0$, as can be seen by considering the unitarity
cuts of the amplitude~\cite{Gaiotto:2011dt}.  These points correspond
to cuts in $u$, $v$ or $w$ originating at either 0 or $\infty$.
A first entry containing $1-u$, $y_u$, {\it etc.},
would lead to a discontinuity starting at an unphysical point.
\end{itemize}

Within our ansatz for the symbol of the three-loop remainder function,
these two constraints are sufficient to show (by explicit enumeration)
that the second entry of the symbol can only be drawn from the set
$\{u,v,w,1-u,1-v,1-w\}$.  This result is consistent with a conjecture
of some of the authors of ref.~\cite{Gaiotto:2011dt}. The second-entry
property is of course true for the known two-loop remainder function.
We also have the following further conditions on the symbol of the
remainder function:
\begin{itemize}
\item It should be completely symmetric in the cross ratios $u,v,w$.
\item It should be parity even.  Because the $y$ variables
of~\eqn{variablesy} invert under parity (the exchange of $z_+$ and
$z_-$), there should be an even number of $y$ entries in any given term
in the symbol.
\item It should vanish in the collinear limit. This constraint
can be implemented at the level of the symbol as follows.
In the limit $w\rightarrow0$, we find that the $y$ variables behave as
\be
       y_u \longrightarrow \frac{u}{1-v} \,, 
\qquad y_v \longrightarrow \frac{v}{1-u} \,,
\qquad y_w \longrightarrow \frac{w(1-u)(1-v)}{(1-u-v)^2}\,.
\label{y_as_wto0}
\ee
The collinear limit can be obtained by first taking the
$w\rightarrow 0$ limit, factorising the symbol and then taking the
limit $v\rightarrow 1-u$. The symbol of the remainder function
should then vanish. (A term in the symbol vanishes if at least one of
its entries goes smoothly to 1.)
\end{itemize}

We have analysed the implications of the above constraints up to three
loops ({\it i.e.}~up to symbols of degree six). At one loop we find that
there are no symbols obeying all of the above properties. This result is
expected, since the remainder function, which vanishes in the collinear
limit, starts appearing only at two loops and beyond.  At two loops there
is a four-parameter family of symbols obeying the constraints that
we have outlined. Not surprisingly, it contains the symbol of the two-loop
remainder function which is explicitly known~\cite{Goncharov:2010jf}
and satisfies the above conditions.  At three loops we find a
59-parameter space of symbols obeying the constraints outlined
above. We would like to impose more constraints to see if we can
further restrict the space of possible solutions. We have the
following two classes of additional constraints:
\begin{itemize}
\item As well as vanishing in the strict collinear limit, the Wilson
loop in the near-collinear regime should have an OPE expansion as
described in refs.~\cite{Alday:2010ku,Gaiotto:2010fk,Gaiotto:2011dt}.
Roughly speaking,
this expansion comes about because a Wilson loop can be expanded
around the limit where a set of adjacent sides becomes collinear.  A
scaling parameter $\tau$ measures how close the Wilson loop is
to the collinear configuration ($\tau \rightarrow \infty$
corresponds to the strict collinear limit). In terms of this parameter
the Wilson loop\footnote{More accurately, one considers the logarithm
of a particular finite, conformally invariant ratio of Wilson loops.}
should have an expansion of the form
\be
W = \int dn \, C_n \, e^{- E_n \tau}\,.
\label{OPEgeneral}
\ee
Here $n$ is shorthand for the set of labels corresponding to the state
being exchanged, $E_n$ is the `energy' of the state ({\it i.e.}~its
eigenvalue under the $\tau$ scaling), and $C_n$ corresponds roughly to
the probability of emission and absorption of a given state. In principle,
a complete knowledge of the set of states labeled by $n$, and
expressions for the energies $E_n$ and the overlap functions $C_n$
entirely fix the remainder function.  In fact, armed with a knowledge
of only the leading corrections to the energies of the simplest
single-particle states, we can predict the {\it leading} discontinuity
at any loop order. At two loops this information is sufficient to
determine the entire symbol~\cite{Gaiotto:2011dt}, because the leading
discontinuity is just a single discontinuity, $\Delta_v R_6^{(2)}$.  
The discontinuities in the other two cross ratios, $\Delta_u R_6^{(2)}$
and $\Delta_w R_6^{(2)}$, are related by symmetry.  Using the fact that
the first entry of the symbol is either $u$, $v$ or $w$, and
\eqn{discsymbol} for the symbol of the discontinuity, we see that knowing
$\Delta_v R_6^{(2)}$ allows the full two-loop symbol to be reconstructed
by appending a $v$ to the front and summing over cyclic permutations.
At three loops, the leading corrections to the $E_n$ suffice to constrain
the double discontinuity, $\Delta_v \Delta_v R_6^{(3)}$.  This is a powerful
constraint, although it does not uniquely determine the remainder function
on its own.
\item The remainder function should also obey particular constraints
in multi-Regge kinematics~\cite{Lipatov:2010qg,Lipatov:2010ad,%
Bartels:2010tx,Bartels:2011xy}.  In this limit, $u \rightarrow 1$,
while $v$ and $w$ vanish in a particular way,
\be\label{multireggekinematics}
u \longrightarrow 1\,, 
\qquad \frac{v}{1-u} \longrightarrow x\,,
\qquad \frac{w}{1-u} \longrightarrow y\,.
\ee
Here $x$ and $y$ are free parameters.\footnote{The variable $y$ introduced
in \eqn{multireggekinematics} should not be confused with the variables
$y_u$, $y_v$ and $y_w$ used for generic kinematics.}
 One must be careful about the
branch on which the limit is taken.  In fact, the functions we are
interested in vanish in this limit in the Euclidean
region~\cite{Brower:2008nm,Brower:2008ia,DelDuca:2008jg}
(when all separations $x_{ij}$ are taken to be spacelike)
but are non-vanishing and even logarithmically divergent
in physical regions for $2 \rightarrow 4$ and $3 \rightarrow 3$
processes~\cite{Bartels:2008ce,Bartels:2008sc,Brower:2008ia,%
Lipatov:2010ad,Bartels:2010tx}.
\end{itemize}

The symbol of the two-loop remainder function is entirely fixed by the
OPE~\cite{Gaiotto:2011dt} to agree with the symbol of the expression
found in ref.~\cite{Goncharov:2010jf}.  The two-loop remainder function
has also been shown to obey the multi-Regge
constraints~\cite{Lipatov:2010qg,Lipatov:2010ad}.
At three loops we find that, of the 59 independent
symbols obeying integrability, symmetry and the collinear limit, 26
have no double discontinuity in a given channel. These functions
therefore cannot be constrained by the OPE analysis.  For the remaining
33 symbols we find that there does exist a {\it unique} solution to the
OPE constraints (thus adding support to the correctness of the ansatz
we have adopted). Thus the OPE fixes 33 of the 59 free parameters of our
symbol.

Analysing the multi-Regge limits we find that, of the 26 functions without
any double discontinuity, only three are non-vanishing in the multi-Regge
kinematics. One has beyond-leading-log behaviour
(it is proportional to $\log^3(1-u)$ in the
limit~(\ref{multireggekinematics})), and is therefore ruled out.
Another parameter is fixed by the known leading-log
behaviour, proportional to $\log^2(1-u)$~\cite{Lipatov:2010ad,Bartels:2010tx}.
Thus a single parameter is left undetermined in the multi-Regge limit.
This free parameter appears in the next-to-leading log behaviour,
but not at the next-to-next-to-leading log level.  We will see later
that it has to be set to zero for consistency with the all-loop-order
prediction concerning $3\to3$ scattering~\cite{Bartels:2010tx}.

Having examined the consequences of the above constraints, we find a
symbol of the form,
\be
\mathcal{S}(R_6^{(3)}) = \mathcal{S}(X) 
+ \sum_{i=1}^{26} \alpha_i \, \mathcal{S}(f_i)\,.
\label{param26}
\ee
The first term, $\mathcal{S}(X)$, is the piece that is fixed by the OPE
constraints. The remaining free parameters $\alpha_i$ accompany
symbols of functions $f_i$ which have no double
discontinuity. Examining the form of $\mathcal{S}(X)$ we find it can
be written in such a way that its final entries are always of the
form, 
\be \Bigl\{ \frac{u}{1-u} , \frac{v}{1-v}, \frac{w}{1-w} , y_u,
y_v, y_w \Bigr\}\,.
\label{finalentries}
\ee
Note that this is not in contradiction with the ansatz~(\ref{ansatz}),
since the entries can always be factorised.  Instead it is a more restrictive
statement, because only 6 out of the 9 potential variables actually appear
in the final entry.  This result concerning the restricted
structure of the final entries of ${\mathcal S}(X)$ is closely connected to
the observations of ref.~\cite{CaronHuot:2011ky}, which suggests that this
fact may be related to a supersymmetric formulation of the Wilson
loop.  Similar restrictions appear~\cite{DDHtoappear}
in differential equations
obeyed~\cite{Drummond:2010cz,Dixon:2011ng,DelDuca:2011wh}
by integrals related to planar $\cN=4$ super-Yang-Mills
scattering amplitudes~\cite{ArkaniHamed:2010kv}.
These observations suggest that the full symbol
$\mathcal{S}(R_6^{(3)})$, not just ${\mathcal S}(X)$,
should be of a form in which its final entries
are drawn from the list~(\ref{finalentries}). Imposing this condition
on the final entries of $\mathcal{S}(R_6^{(3)})$ reduces the number
of free parameters to just two. The fact that it is possible to impose
this restriction, consistently with the known multi-Regge behaviour,
is highly non-trivial.

Finally, let us note that even if we were able to fix the entire symbol
and find a function with all the desired analytic properties,
vanishing in the collinear limit with the correct OPE behaviour, {\it etc.},
there would always remain the possibility of adding some amount of the
two-loop remainder function multiplied by $\zeta_2$, that is,
\be
R_6^{(3)}\ \longrightarrow\ R_6^{(3)} + \gamma \, \zeta_2 \, R_6^{(2)}\,,
\label{shiftambig}
\ee
for some constant $\gamma$.
We will see such `beyond-the-symbol' ambiguities appearing in a
particular way in our predictions for the multi-Regge behaviour of the
three-loop remainder function.

We will now discuss the OPE analysis in further detail, and then
describe the predictions for the three-loop remainder function in the
multi-Regge kinematics.  We will conclude with a discussion of
the conditions on the final entries, and the remaining ambiguities
after imposing all our constraints.


\section{OPE Constraints}
\label{sect-OPE-constraints}

\begin{figure}
\centerline{{\epsfysize4.5cm \epsfbox{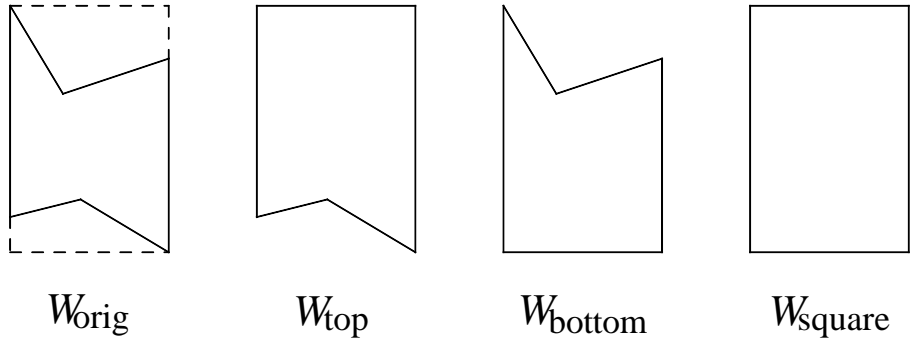}}} \caption[]{\small
The four different Wilson loops entering the definition of the
ratio~(\ref{logratio}). The reference square is shown by the dashed
line.  The top and bottom Wilson loops are obtained by replacing a
sequence of edges by the corresponding part of the square.}
  \label{Wloops}
\end{figure}

In order to describe the OPE expansion for a light-like Wilson loop,
the authors of ref.~\cite{Alday:2010ku} introduced a reference square
with null sides, denoted by $W_{\rm square}$ in Fig.~\ref{Wloops}.
Two of the sides of the square coincide with two of
the sides of the Wilson loop, while the other two sides are formed by
finding other null lines that intersect the two previous ones as well
as two of the corners of the original loop. One can then consider the
finite, conformally invariant quantity made from a ratio of Wilson
loops,
\be r = \log \frac{W_{\rm orig} W_{\rm square}}{W_{\rm top}
W_{\rm bottom}}\,.
\label{logratio}
\ee
The four different Wilson loops appearing in the ratio are depicted in
Fig.~\ref{Wloops}.  

Note that at six points, the top and bottom loops are five-sided.
The four-sided and five-sided loops appearing in the ratio $r$ are
entirely determined by the conformal Ward identity~(\ref{confWI}).
Thus knowledge of the quantity $r$ is equivalent to knowing the
six-point remainder function.

As described in ref.~\cite{Gaiotto:2011dt}, the Wilson loop, 
or more precisely the ratio $r$, is expected to have an OPE expansion
of the form
\be
r = \int dn \, C_n \, e^{- E_n \tau}\,.
\label{OPEgeneral2}
\ee
At one loop, the states labelled by $n$ are free single-particle
exchanges between the bottom half of the the loop and the top
half. Beyond one loop there can be interactions between the particles
and the vertical Wilson lines in Fig.~\ref{Wloops},
as well as multi-particle exchanges, and so forth. 
The quantities $C_n$ and $E_n$ entering the OPE should be expanded in
the coupling constant.  In principle, to determine $r$
(and hence the remainder function) one needs to know the space of
states and the dependence of $C_n$ and $E_n$ on the coupling.

There is, however, a piece of the remainder function that is completely
constrained at $l$ loops, just from knowing the one-loop anomalous
dimensions~\cite{Basso:2010in} of the single-particle states being
exchanged~\cite{Gaiotto:2011dt}.  In the near-collinear limit, one of
the cross ratios vanishes, say $v\to0$.  It vanishes exponentially
quickly as $\tau\to\infty$; that is, $\tau$ is proportional to $\ln v$
in this limit.
The special piece of the remainder function (or $r$) is the leading
discontinuity in $v$, which is the repeated $(l-1)$-fold
discontinuity $\Delta_v^{l-1} r$.  This discontinuity can be extracted
from the OPE by first Taylor expanding the energies of the intermediate
states in the coupling constant,
\be
E_n = E_n^{(0)} + g^2 E_n^{(1)} + g^4 E_n^{(2)} + \ldots\,.
\ee 
After Taylor expanding the exponential in~\eqn{OPEgeneral} in $g^2$
we find
\be r = \int dn \, C_n \, e^{- E_n^{(0)}\tau}
\Bigl[1 - g^2 \tau E_n^{(1)} 
+ g^4 \bigl( \tfrac{1}{2} \tau^2 (E_n^{(1)})^2 - \tau E_n^{(2)} \bigr)
 + \ldots \Bigr] \,.
\label{ExpandExp}
\ee
Because $\tau$ is proportional to $\ln v$ as $\tau\to\infty$,
the leading discontinuity in $v$
at any loop order is given by the term involving the highest power of $\tau$.
This term is always obtained from the one-loop corrections $E_n^{(1)}$
to the energies of the simplest single-particle states --- those states
whose overlap functions $C_n$ are non-vanishing at order $g^2$.

The exchanged states carry other quantum numbers in addition to the
energy $E_n$. There is a `momentum' $p$ conjugate to the other scaling
($\sigma$-scaling) invariance of the square and a discrete label $m$,
conjugate to the rotational invariance ($\phi$-rotation) in the two
directions orthogonal to the square. These three invariances of the
square can be used to completely parametrise the three variables $u,v$
and $w$ on which $r$ (or the six-point remainder function)
depends. Explicitly, the variables $\sigma, \tau$ and $\phi$ are
related to $u,v$ and $w$ via
\be
u = \frac{e^\sigma \sinh\tau \tanh\tau}
   {2 (\cosh\sigma \cosh\tau + \cos\phi)}\,, 
\quad v=\frac{1}{\cosh^2\tau}\,, 
\quad w= \frac{e^{-\sigma} \sinh\tau \tanh\tau}
{2 (\cosh\sigma \cosh\tau + \cos\phi)}\,.
\label{uvwtausigmaphi}
\ee

A more detailed description of the leading discontinuity of $r$ at $l$
loops is then
\bea
\Delta_v^{l-1} \, r^{(l)} \ &\propto&
\frac{(-1)^l}{(l-1) !} \int \frac{dp}{2\pi} e^{-ip\sigma} 
\, \biggl(\sum_{m=1}^{\infty} 
\frac{[\gamma_{m+2}(p)]^{l-1} \cos (m \phi)}{p^2 + m^2}
\nonumber\\
&&\hskip3.3cm\null
+ \sum_{m=2}^{\infty} \frac{[\gamma_{m-2}(p)]^{l-1} 
  \cos( (m-2)\phi)}{p^2 + (m-2)^2}\biggr)
 \mathcal{C}_m (p) \mathcal{F}_{m/2,p}(\tau)\,.
\label{l-loopdisc}
\eea
The one-loop anomalous dimensions $\gamma_m(p)$ are the energies
$E_n^{(1)}$ of conformal primary states, and they are given
by~\cite{Basso:2010in},
\be
\gamma_m(p) = 
  \psi\bigl(\tfrac{m+ip}{2}\bigr)
+ \psi\bigl(\tfrac{m-ip}{2}\bigr)
- 2 \, \psi(1) \,.
\label{gammamp}
\ee
The explicit formulae for the overlap functions $\mathcal{C}_m(p)$ 
and the conformal blocks $\mathcal{F}_{m/2,p}(\tau)$, which account
for the exchange of conformal descendant states, are given in
ref.~\cite{Gaiotto:2011dt}.
The formula~(\ref{l-loopdisc}) has been slightly adapted from the
corresponding one for two loops~\cite{Gaiotto:2011dt} by raising
the anomalous dimensions $\gamma_m(p)$ to the power $l-1$
instead of 1.  This power originates simply from the highest
power of $\tau$ at each loop order in \eqn{ExpandExp}, as in this term
the anomalous dimension appears in the exponent accompanied by a
factor of $\tau$.  In summary, the leading discontinuity in any of the
cross ratios (which are all related by the permutation symmetry)
is completely predicted by the OPE, in a formula very similar to the
two-loop case.

Evaluating the expression~(\ref{l-loopdisc}) is quite
involved.  However, following ref.~\cite{Gaiotto:2011dt} we can say
that it must obey certain differential equations.  The differential
operators $\mathcal{D}_\pm$ of ref.~\cite{Gaiotto:2011dt} should
annihilate any function given by a sum of two towers of conformal
blocks.  Using results from Appendix A, one can work out the form
of these operators in terms of the cross ratios $u$, $v$ and $w$:
\begin{align}
\mathcal{D}_\pm = \frac{4}{1-v}\Big[ 
&- z_\pm u \partial_u -(1-v)v \partial_v - z_\pm w \partial_w \notag \\
&+ (1-u)v u\partial_u u \partial_u 
 + (1-v)^2 v \partial_v v \partial_v
 + (1-w)vw\partial_w w \partial_w \notag \\
&+ (-1+u-v+w)\bigl((1-v)u\partial_uv\partial_v
 - vu \partial_u w \partial_w + (1-v)v\partial_v w \partial_w\bigr)\Bigr]\,.
\label{Dpmuvw}
\end{align}

At any given loop order beyond one loop, the symbol of the remainder
function $R_6^{(l)}$ is actually equal to the symbol
of the Wilson loop ratio $r^{(l)}$. The difference between the two functions
comes from additional terms in the expansion of \eqn{logratio} in the
coupling.  For example, in \eqn{WardIdSoln}, $R_n$ is a constant for
the four- and five-point contributions to $r$, but there are degree
two functions (at most) related to $F^{\rm WL}_{n\,,{\rm 1-loop}}$
that will contribute to the difference between $r$ and the remainder
function, when they are multiplied by transcendental constants from
higher-order terms in $\gamma_K$.  These terms drop out of
the symbol.

For our three-loop analysis we require that the symbol of the
leading (double) discontinuity $(\Delta_v)^2 R_6^{(3)}$ is annihilated
by the product of $\mathcal{D}_+$ and $\mathcal{D}_-$,
\be
\mathcal{S}\bigl(\mathcal{D}_+ \mathcal{D}_-
\Delta_v \Delta_v R_6^{(3)}(u,v,w)\bigr) = 0\,.
\label{DpDm=0}
\ee
The above is a very general constraint, which should apply to all
expressions admitting an OPE expansion of the form described in
ref.~\cite{Alday:2010ku}. Within our specific ansatz it becomes
extremely powerful.  We find that it fixes 33 out of the 59
coefficients that were undetermined after imposing integrability,
symmetry and the collinear limit.  The remaining 26 terms have no
double discontinuity in any single channel, so they cannot be fixed
without supplying additional information.

In ref.~\cite{Gaiotto:2011dt} the sum~(\ref{l-loopdisc}) was performed
for the single discontinuity $\Delta_v R_6^{(2)}$ at two loops, for which
only a single power of the anomalous dimensions $\gamma_m(p)$
appears. One method used is to compute the discontinuities of the
discontinuity. We can perform a similar analysis for the double
discontinuities of the double discontinuity at three loops.

At two loops the discontinuity $\Delta_v R_6^{(2)}$ has further
discontinuities of the type $\Delta_u$, $\Delta_w$ and $\Delta_{1-v}$.
The double discontinuity $\Delta_w \Delta_v R_6^{(2)}$ is a degree two
function.  When computing the integral~(\ref{l-loopdisc}) over $p$ as a
sum over residues, it arises from double poles in the $p$ plane for $l=2$.
The formula~(\ref{gammamp}) for $\gamma_m(p)$ contains single poles, with
constant residues, at $p = i (m+2k)$ for non-negative integers $k$.  They
can combine with single poles at the same locations in the overlap
functions $\mathcal{C}_m(p)$.  For $p=im$ they can also combine
with poles from the $p^2+m^2$ denominator factor.
Double poles give rise to derivatives with respect to $p$, which can
hit the exponential $e^{-ip\sigma}$ (the only place $\sigma$ appears)
and bring down a factor of $\sigma$.  Because $\ln w$ is proportional
to $\sigma$ in \eqn{uvwtausigmaphi} as $\sigma \to +\infty$, 
the coefficient of the term linear in $\sigma$ yields the discontinuity
with respect to $w$.

Similarly, at three loops the double discontinuity of the double
discontinuity $\Delta_w \Delta_w \Delta_v \Delta_v R_6^{(3)}$ arises
from triple poles in the $p$ plane in the expression~(\ref{l-loopdisc})
for $l=3$, which generate two derivatives with respect to $p$ acting
on $e^{-ip\sigma}$. The analysis of appendix B.1 of
ref.~\cite{Gaiotto:2011dt} is almost directly applicable to
$\Delta_w \Delta_w \Delta_v \Delta_v R_6^{(3)}$.
However, there is a small mismatch due to the factor of
$p^2 + m^2$ in the denominator of the terms in the first sum
in~\eqn{l-loopdisc}. This factor contributes a pole at $p=im$,
which combines with the pole coming from $\mathcal{C}_m(p)$ to
produce a double-pole contribution to the two-loop expression
$\Delta_w \Delta_v R_6^{(2)}$, without requiring a pole from
$\gamma_{m+2}(p)$.  There are no such contributions for the three-loop
expression $\Delta_w \Delta_w \Delta_v \Delta_v R_6^{(3)}$, because the
only triple poles come from combining $[\gamma_m(p)]^2$ with
$\mathcal{C}_m(p)$. 

On the other hand, if we could remove the $p^2 + m^2$ factor in the
denominator of \eqn{l-loopdisc}, then the same analysis for the two-loop
problem would also apply directly at three loops.  It is important for
this conclusion that the residues of $\gamma_m(p)$ are constants,
independent of $m$ and $p$.
Removing the denominator amounts to acting with the particular
second-order operator $\Box = -(\partial_\sigma^2 + \partial_\phi^2)$
described in ref.~\cite{Gaiotto:2011dt}.
In terms of the cross ratios, using results from Appendix A,
the operator $\Box$ is given by
\be
\Box = \frac{4 u w}{1-v}
\Bigl[ u \partial_u + w \partial_w - (1-u)\partial_u u \partial_u
    - (1-w)\partial_w w \partial_w 
    + (1-u-v-w + 2 uw)\partial_u \partial_w\Bigr]\,.
\label{boxop}
\ee
We therefore conclude that
\be
 \Box \Delta_w \Delta_w \Delta_v \Delta_v R_6^{(3)}
 \ \propto\ \Box \Delta_w \Delta_v R_6^{(2)}
 = \frac{w(1-u+v-w)}{(1-v)(1-w)}\,.
\label{boxconstraint}
\ee
The second equation can be found by acting with $\Box$ on the
symbol for the discontinuity of $R_6^{(2)}$,
\bea
\mathcal{S}(\Delta_w \Delta_v R_6^{(2)}) &=& - \frac{1}{4} \Bigl\{
 u \otimes \frac{uvw}{(1-u)(1-v)(1-w)} \, 
- \, (1-w) \otimes \frac{v}{1-v} \, - \, (1-v) \otimes \frac{w}{1-w} 
\nonumber\\ &&\hskip0.7cm\null
- y_u \otimes y_u y_v y_w \Bigr\} \,,
\label{dwdvR62symb}
\eea
which is easily extracted from the symbol~(\ref{R62symb})
for $R_6^{(2)}$.  It can also be found by applying $\Box$ to the
explicit representation for the discontinuity $X_3$ found in
ref.~\cite{Gaiotto:2011dt}.
(We have not yet fixed the overall normalization of $R_6^{(3)}$;
we will do this subsequently when we match to the leading-logarithmic
behaviour in the multi-Regge limit.)
Remarkably, the symbol obtained after imposing the
condition~(\ref{DpDm=0}) is perfectly consistent with the
condition~(\ref{boxconstraint}), which is a non-trivial check
of our analysis.

In conclusion, after imposing the leading OPE constraints we find a
solution consistent with our ansatz containing 26 unfixed parameters
$\alpha_i$,
\be
\mathcal{S}(R_6^{(3)})
= \mathcal{S}(X) + \sum_{i=1}^{26} \alpha_i \, \mathcal{S}(f_i)\,.
\label{symbolafterOPE}
\ee
Each of the symbols appearing in the above expression is required
to be integrable, and so there do exist functions $X, f_i$ with
those symbols.  The
double discontinuities of $X$ and the $f_i$ obey
\be
\Delta_v \Delta_v X \neq  0, 
\qquad \mathcal{D}_+ \mathcal{D}_- \Delta_v \Delta_v X = 0,
\qquad \mathcal{S}(\Delta_v \Delta_v f_i) =0\,.
\ee
Although the symbol for $X$ is one of the central results of this
article, it is also rather lengthy.  Therefore we do not present it
directly in the text.  Instead we give it in accompanying Mathematica
and plain text files.  In these files, a term
$a\otimes b\otimes \ldots \otimes f$ is written as SB$[a,b,\ldots,f]$.
Using symbol(ic) manipulation programs, it is
straightforward to extract information about various limits and
discontinuities from the symbol.  The next section describes one such limit,
multi-Regge kinematics.


\section{Predictions for multi-Regge kinematics}
\label{sect-Regge-predictions}

We now analyse our symbol in the multi-Regge
limit~(\ref{multireggekinematics}), in which $u\to1$ while $v$ and $w$
vanish.
First we find that in the Euclidean version of this limit, the symbol we
have found vanishes, in agreement with
observations~\cite{Brower:2008nm,Brower:2008ia,DelDuca:2008jg}
about the consistency of the BDS ansatz in this type of limit.
Next we analytically continue to a physical branch, defined by letting
$u \rightarrow e^{-2\pi i} u$.  For physical $2\to4$ scattering, $v$ and $w$
remain at their Euclidean values. The imaginary terms on the physical branch
that are generated by this transformation of $u$ come from
the discontinuity of the function in the $u$ channel in
the multi-Regge limit.  
As mentioned in section~\ref{sect-symbols}, 
the symbol of the discontinuity of a function $f$ in a given channel
($u$ say) can be found by taking the terms in the original symbol
$\mathcal{S}(f)$ with initial entry $u$ and stripping off that entry.
The result, after multiplying by $(-2 \pi i)$, is the symbol of the
discontinuity $\mathcal{S}(\Delta_u f)$.  The real terms for $2\to4$
scattering come from a double discontinuity in the $u$ channel.
They are found from $\mathcal{S}(\Delta_u \Delta_u f)$, after
multiplying by $(2 \pi i)^2$.  (In principle, there can be contributions
to the imaginary and real parts from triple and higher order
discontinuities in $u$ as well. However, through three loops there
are no such terms.)

The behaviour we expect for the $l$-loop remainder function in the
multi-Regge limit in the physical region is
\be
R_6^{(l)}\ \longrightarrow\
(2\pi i) \sum_{r=0}^{l-1} \log^r(1-u) \, 
\Bigl[ g^{(l)}_r(x,y) \, + \, 2\pi i \, h^{(l)}_r(x,y) \Bigr] \,,
\label{R6lregge}
\ee
where the logarithmic expansion coefficients $g^{(l)}_r$ and $h^{(l)}_r$
are functions that depend only on the finite ratios $x$ and $y$ defined in
\eqn{multireggekinematics}.
It is convenient to change variables to describe these functions.
Following ref.~\cite{Lipatov:2010ad}, we introduce the
variables $w,\ws$ defined by\footnote{The new variable $w$ in
the multi-Regge limit (which is always accompanied by a $w^*$)
should not be confused with the original cross ratio $w$.}
\be
x = \frac{1}{(1+w)(1+\ws)} \,, \qquad y = \frac{w \ws}{(1+w)(1+\ws)} \,.
\label{wwsdef}
\ee
In terms of these variables, the symbols of the functions
$g_r^{(l)}$ and $h_r^{(l)}$
have as their only entries $w,\ws$, $(1+w)$, and $(1+\ws)$.

Both $g_r^{(l)}$ and $h_r^{(l)}$ are invariant under two $Z_2$ symmetries:
\be
{\rm conjugation:}\quad w \longleftrightarrow \ws,
\label{conjugation}
\ee
which is a reality condition for the case that $\ws$ is the
complex conjugate of $w$, and
\be
{\rm inversion:}\quad  w \longleftrightarrow 1/w, \qquad
                \ws \longleftrightarrow 1/\ws.
\label{inversion}
\ee
The combined operation of inversion and conjugation is the reflection 
symmetry $x\leftrightarrow y$, which is inherited from the permutation
symmetry $v\leftrightarrow w$ for generic kinematics.
We also expect the functions $g_r^{(l)}$ and $h_r^{(l)}$ to be
single-valued as $w$ is rotated around the origin of the complex
plane.  Finally, the functions should vanish for $|w| \rightarrow 0$,
which is the collinear limit on top of the Regge limit.

In taking the multi-Regge limit~(\ref{multireggekinematics}) of symbols,
we note that any symbol containing a $u$ or a $y_u$ entry can be
discarded, because $u\to1$ and $y_u\to1$ in this limit.
We recall the definition of $x$ and $y$ in~\eqn{multireggekinematics}.
The variables $y_v$ and $y_w$ go to finite values,
$\tilde{y}_v$ and $\tilde{y}_w$, in the limit:
\bea
y_v &\longrightarrow& \tilde{y}_v 
= \frac{-1 -x +y + \sqrt{\tilde{\Delta}}}{-1-x+y-\sqrt{\tilde{\Delta}}}
= \frac{1+w^*}{1+w}  \,,
\label{yvtilde}\\
y_w &\longrightarrow& \tilde{y}_w
= \frac{-1+x-y + \sqrt{\tilde{\Delta}}}{-1+x-y-\sqrt{\tilde{\Delta}}}
= \frac{(1+w)w^*}{w(1+w^*)}  \,,
\label{ywtilde}
\eea
where $\tilde{\Delta} = (1-x-y)^2 - 4xy$ is the limit of
$\Delta/(1-u)^2$.  The relation of $\ty_v$ and $\ty_w$ 
to the $(w,w^*)$ variables can be found with the aid of
formulae in Appendix B.

The symbols $\mathcal{S}(g_r^{(l)})$ and $\mathcal{S}(h_r^{(l)})$
do not fix the functions $g_r^{(l)}$ and $h_r^{(l)}$ uniquely.
One can always add transcendental constants such as $\zeta_2$,
multiplied by lower transcendentality functions
which vanish in the symbol.  However, the above symmetries,
\eqns{conjugation}{inversion}, and analytic properties around $w=0$,
greatly restrict the form of such potential ambiguities. In particular
there are no such functions of degree 0 or 1 obeying these constraints.

Before describing the three-loop predictions, we
recall~\cite{Lipatov:2010qg,Lipatov:2010ad}
the corresponding expansion~(\ref{R6lregge})
at two loops, as obtained from the formula of
ref.~\cite{Goncharov:2010jf},
\bea
g_1^{(2)}(w,\ws) &=& \frac{1}{4} \, \ln|1+w|^2 \, \ln \frac{|1+w|^2}{|w|^2} \,,
\label{g(2)1}\\
g_0^{(2)}(w,\ws) &=& \frac{1}{4} \, \ln|w|^2 \, \ln^2|1+w|^2
  - \frac{1}{6} \, \ln^3|1+w|^2
  + \frac{1}{2}\ln|w|^2 \Bigl[ \polylog_{2}(-w) + \polylog_{2}(-\ws) \Bigr]
\nonumber\\
&&\hskip0.0cm\null
  - \polylog_{3}(-w) - \polylog_{3}(-\ws) \,.
\label{g(2)0}
\eea
It is not always the case that $\ws$ is the complex conjugate of $w$.
(That only happens if $\sqrt{\tilde\Delta}$ is imaginary.)  In the
general case, $|w|^2$ is just a shorthand for $w\ws$, and $|1+w|^2$ is
a shorthand for $(1+w)(1+\ws)$.

The functions controlling the real parts depend on whether the scattering
is $2\to4$ or $3\to3$.  In $2\to4$ scattering, the multi-Regge limit
has vanishing real part at two loops~\cite{Lipatov:2010qg},
\bea
h_1^{(2)}(w,\ws) &=& 0 \,,
\label{h(2)1}\\
h_0^{(2)}(w,\ws) &=& 0 \,.
\label{h(2)0}
\eea
In the case of $3\to3$ scattering, $v$ and $w$ have to be analytically
continued to the opposite sign from their Euclidean
values~\cite{Bartels:2010tx}; that is,
\be
u \to |u| e^{2\pi i} \,, \qquad
v \to |v| e^{\pi i} \,, \qquad
w \to |w| e^{\pi i} \,.
\ee
In fact, the remainder function for $3\to3$ scattering can be derived
from the $2\to4$ case by the simple substitution
\be
\log(1-u)\ \longrightarrow\ \log(u-1) - i \pi \,,
\label{lnomushift}
\ee
followed by complex conjugation~\cite{Bartels:2010tx}.

Whereas the function $g^{(2)}_1$ in~\eqn{g(2)1} is manifestly invariant
under both conjugation and inversion symmetries, $g^{(2)}_0$
in~\eqn{g(2)0} only has manifest invariance under 
$w\leftrightarrow \ws$.  On the other hand, this form makes clear
that $g^{(2)}_0$ vanishes as $|w| \rightarrow 0$, and also that
it acquires no phase as $w$ is rotated around the origin of the
complex plane. The latter property is obvious for $|w|<1$ and true by
inversion symmetry for $|w|>1$.  Simple polylogarithm identities can
be used to demonstrate the $w$ inversion symmetry. In fact, assuming
maximal transcendentality, the functions $g_1^{(2)}$ and $g_0^{(2)}$,
of degree 2 and 3 respectively, can be fixed uniquely,
just by knowing the symbol of the two-loop
remainder function and imposing these requirements.  The uniqueness holds
because the symbol fixes the functions up to constants like $\zeta_3$ or
$\zeta_2$, multiplied by functions of corresponding lower degree,
and there are no functions with degree 0 or 1 obeying the constraints.

At three loops we find that in the multi-Regge limit,
the symbol $\mathcal{S}(X)$ has the form of the symbol of the right-hand
side of~\eqn{R6lregge} for $l=3$,
with the leading divergence being a double logarithmic one.  We also
find that in this limit, all but three of the $\mathcal{S}(f_i)$
vanish. We will call the functions with non-vanishing symbols in the
limit $f_{24},f_{25},f_{26}$. We find that one symbol,
$\mathcal{S}(f_{26})$, has a triple logarithmic divergence in the
multi-Regge limit, which is one logarithm beyond the known degree of
divergence. Therefore the coefficient $\alpha_{26}$ must vanish. The
symbol $\mathcal{S}(X)$ contributes to the double logarithmic
divergence exactly what is required to match the symbol of the
leading-log prediction~\cite{Lipatov:2010ad}. We find that
$\mathcal{S}(f_{25})$ also contributes a double logarithmic divergence
(different in form from that of $\mathcal{S}(X)$).  Hence we deduce that its
coefficient $\alpha_{25}$ must vanish, so that it does not spoil the
agreement with the leading-log prediction. The final symbol
$\mathcal{S}(f_{24})$ then contributes to the next-to-leading-log term
({\it i.e.}~to $\mathcal{S}(g^{(3)}_1)$) but not to the
next-to-next-to-leading one ({\it i.e.}~not to $\mathcal{S}(g^{(3)}_0)$).
Because it is the only arbitrary
coefficient from the expression~(\ref{symbolafterOPE}) that survives
in the multi-Regge limit (after imposing the constraints we have
just discussed), we give it a new name, $\alpha_{24} = c$.

Now we describe our predictions for the multi-Regge limit, 
after imposing the conditions,
\be
\alpha_{24} = c, \qquad \alpha_{25} = 0, \qquad \alpha_{26}=0\,.
\label{reggeconstraints}
\ee
We find (as described above) that the symbol of $g^{(3)}_2$ agrees
precisely with the symbol of the coefficient of the $\log^2(1-u)$ term
predicted in ref.~\cite{Lipatov:2010ad}, namely
\be
\mathcal{S}( g^{(3)}_2 ) 
= \frac{1}{32}\Bigl( 2 \, x \otimes x \otimes y 
            + 3 \, x \otimes y \otimes xy
            - x \otimes \tilde{y}_w \otimes \tilde{y}_v\tilde{y}_w \Bigr)
            \ +\ (x \longleftrightarrow y)\,.
\label{symbolg(3)2}
\ee
We have adjusted the overall normalization of $X$ so that this term
in the multi-Regge limit agrees with ref.~\cite{Lipatov:2010ad}.
This normalization is based on the loop expansion parameter
$a = g^2 N_c/(8\pi^2)$ and \eqn{Rnnorm}.

When written in terms of the $w,w^*$ variables, the
symbol~(\ref{symbolg(3)2}) can be seen to be the symbol of the
following function,
\be
g^{(3)}_2(w,\ws) = \frac{1}{8} \, g^{(2)}_0(w,\ws)
- \frac{1}{32} \, \ln|1+w|^2 \, \ln\frac{|1+w|^2}{|w|^2} 
\, \ln\frac{|1+w|^4}{|w|^2} \,,
\label{g2wto0}
\ee
exactly as predicted in ref.~\cite{Lipatov:2010ad}. Just as in the
two-loop case, this degree 3 function is uniquely determined by its symbol,
because there are no suitable degree 1 or 0 functions one could add.
Also, we find from the double $u$ discontinuity that the real part
at leading-log level vanishes,
\be
h^{(3)}_2(w,\ws) = 0 \,,
\label{h2wto0}
\ee
as expected.

We also have predictions for the symbols of $g^{(3)}_1$,
$g^{(3)}_0$ and $h^{(3)}_0$ (and their corresponding functions)
which are new.  The function $h^{(3)}_1$ for $2\to4$
kinematics was predicted in ref.~\cite{Lipatov:2010ad}, and we
obtain the same function. Remarkably, all these functions can be
expressed in terms of classical polylogarithms.

As the transcendental degree increases, it becomes more difficult to
write the result in a form that is simultaneously invariant under
inversion of $w$, and manifestly has good behaviour as 
$|w| \to 0$. We choose to express the functions in a form where
the $|w| \to 0$ behaviour is manifest.  (Alternate forms with manifest
inversion symmetry can be found in Appendix C.)  At the
next-to-leading-log level, we find
\bea
g^{(3)}_1(w,\ws) &=& \frac{1}{8} \Biggl\{
 \ln |w|^2 \, 
\biggl[ \polylog_3\left(\frac{w}{1+w}\right)
      + \polylog_3\left(\frac{\ws}{1+\ws}\right) \biggr]
\nonumber\\
&&\hskip-2.5cm\null
+ ( 5 \, \ln|1+w|^2 - 2 \, \ln|w|^2 ) 
  \Bigl[ \polylog_3(-w) + \polylog_3(-\ws) \Bigr] 
\nonumber\\
&&\hskip-2.5cm\null
- \frac{3}{2} \ln|w|^2 \, \ln\frac{|1+w|^4}{|w|^2}
  \Bigl[ \polylog_2(-w) + \polylog_2(-\ws) \Bigr] 
\nonumber\\
&&\hskip-2.5cm\null
- \frac{1}{12} \,  \ln^2|1+w|^2
    \biggl[ \ln|w|^2 \, ( \ln|w|^2 + 2 \, \ln|1+w|^2 )
          - 10 \, \ln^2\frac{|1+w|^2}{|w|^2} \biggr]
\nonumber\\
&&\hskip-2.5cm\null
+ \frac{1}{2} \, \ln|w|^2 
 \, \ln\frac{|1+w|^2}{|w|^2} \, \ln(1+w) \, \ln(1+\ws)
- 2 \, \zeta_3 \, \ln{|1+w|^2}  \Biggr\}
\nonumber\\
&&\hskip-2.5cm\null
+ \biggl(\frac{5}{2} + \gamma'\biggr) \, \zeta_2 \, g^{(2)}_1(w,\ws)
+ \, c \, g_1^a \,.
\label{g1wto0}
\eea
For this degree-four function there are only two constants to determine.
The first one, $\gamma'$, corresponds to the freedom
to add the two-loop remainder function,
multiplied by $\zeta_2$, to the three-loop remainder function,
as in~\eqn{shiftambig}. 
The second constant, $c$, is the remaining ambiguity at the level
of the symbol.  It multiplies the function,
\bea
g_1^a(w,\ws) &=&
4 \, \ln\frac{|1+w|^2}{|w|^2} 
 \Bigl[ \polylog_3(-w) + \polylog_3(-\ws) \Bigr] 
- 4 \, \ln|w|^2 \, 
\biggl[ \polylog_3\left(\frac{w}{1+w}\right)
      + \polylog_3\left(\frac{\ws}{1+\ws}\right) \biggr]
\nonumber\\
&&\hskip-2.5cm\null
+ 2 \, \Bigl[  \polylog_2(-w) - \polylog_2(-\ws)
    + \ln|w|^2 \, \ln\frac{1+w}{1+\ws} \Bigr]
       \Bigl[  \polylog_2(-w) - \polylog_2(-\ws) \Bigr]
\nonumber\\
&&\hskip-2.5cm\null
+ \frac{1}{6} \, \ln^3|1+w|^2 \, ( \ln|1+w|^2 + 2 \, \ln|w|^2 )
- 2 \, \ln|w|^2 \, \ln|1+w|^2 \, \ln(1+w) \, \ln(1+\ws)
\nonumber\\
&&\hskip-2.5cm\null
+ 8 \, \zeta_3 \, \ln|1+w|^2 \,.
\label{g1awto0}
\eea
We will see later that this function does not enter, {\it i.e.}~that $c=0$,
if we impose consistency with the all-loop-order prediction for $3\to3$
kinematics~\cite{Bartels:2010tx}.
Also, in section~\ref{finalentry} we will see that this function can also
be excluded by imposing an additional constraint on the form of the
final entries in the symbol of $R_6^{(3)}$.

We rule out additional constants multiplying lower-degree
transcendental functions in~\eqn{g1wto0} by first assuming that
potential functions must be built from logarithms and (at high enough
degree) polylogarithms containing the same arguments found in the
leading-transcendentality (symbol-level) terms, namely $\ln w$,
$\ln(1+w)$, $\polylog_m(-w)$, $\polylog_m(w/(1+w))$ and
$\polylog_m(1/(1+w))$ (for $m=2,3$ these polylogarithms are not all
independent).  After enumerating such functions, we impose the four
constraints discussed above: the conjugation and inversion symmetries;
vanishing of the function in the (collinear-Regge)
$|w| \rightarrow 0$ limit; and absence of a phase as $w$ is rotated
around the origin of the complex plane. These constraints rule out
functions of degree 0 or 1. The unique function of degree 2 obeying these
conditions is $g^{(2)}_1(w,w^*)$\,.  If we had omitted the final-entry
condition, for example, we could have added a term proportional to
\be
\zeta_2 \, \ln\left(\frac{1+w}{1+\ws}\right) 
        \, \ln\left(\frac{(1+w)\ws}{(1+\ws)w}\right) \,.
\ee
This term has both symmetries and vanishes as $|w| \rightarrow 0$;
in fact, it is the unique term at degree two that satisfies the
other three constraints but violates the phase condition.

The degree-three function controlling the real part at
next-to-leading-log level, $h^{(3)}_1$, can be found from the
multi-Regge limit of the double $u$ discontinuity.  (There is an overall
factor of $1/2$ associated with the fact that the symbol of $\ln^2 u$
is $2\, u \otimes u$.)  We find
\be
\mathcal{S}(h^{(3)}_1) = \mathcal{S}(g^{(3)}_2)
- \frac{1}{8} \Bigl[
  x \otimes y \otimes y + y \otimes x \otimes y 
+ y \otimes y \otimes x\ +\ (x\longleftrightarrow y) \Bigr] \,,
\label{h3_1_symb}
\ee
which integrates to 
\be
h^{(3)}_1(w,\ws) = g^{(3)}_2(w,\ws)
+ \frac{1}{16} \, \ln |1+w|^2 \, \ln \frac{|1+w|^2}{|w|^2}
               \, \ln \frac{|1+w|^4}{|w|^2} \,.
\label{h1}
\ee
This result agrees with that found in ref.~\cite{Lipatov:2010ad}.

Moving on to next-to-next-to-leading-log level, we find the
degree-five function controlling the imaginary part,
\bea
g^{(3)}_0(w,\ws) &=& -\frac{1}{32} \Biggl\{
- 60 \, \Bigl[ 2 \, \Bigl( \polylog_5(-w) + \polylog_5(-\ws) \Bigr)
    - \ln|w|^2 \, \Bigl( \polylog_4(-w) + \polylog_4(-\ws) \Bigr) \Bigr]
\nonumber\\
&&\hskip-2.5cm\null
+ 12 \, \biggl[ 
  2 \, \biggl( \polylog_5\left(\frac{w}{1+w}\right)
             + \polylog_5\left(\frac{1}{1+w}\right)
             + \frac{1}{24} \, \ln w \, \ln^4(1+w)
\nonumber\\
&&\hskip-1.3cm\null
             + \polylog_5\left(\frac{\ws}{1+\ws}\right)
             + \polylog_5\left(\frac{1}{1+\ws}\right)
             + \frac{1}{24} \, \ln \ws \, \ln^4(1+\ws)
       \biggr)
\nonumber\\
&&\hskip-1.7cm\null
 + \ln\frac{|1+w|^2}{|w|^2} \biggl(
      \polylog_4\left(\frac{w}{1+w}\right)
    + \polylog_4\left(\frac{\ws}{1+\ws}\right) \biggr)
\nonumber\\
&&\hskip-1.7cm\null
 + \ln|1+w|^2 \biggl( \polylog_4\left(\frac{1}{1+w}\right)
             - \frac{1}{6} \, \ln w \, \ln^3(1+w)
\nonumber\\
&&\hskip0.5cm\null
             + \polylog_4\left(\frac{1}{1+\ws}\right)
             - \frac{1}{6} \, \ln \ws \, \ln^3(1+\ws) \biggr) \biggr]
\nonumber\\
&&\hskip-2.5cm\null
- 2 \, \Bigl( 5 \, ( \ln^2|w|^2 - \ln^2|1+w|^2 )
            + 6 \, \ln|w|^2 \, \ln|1+w|^2 \Bigr)
  \Bigl( \polylog_3(-w) + \polylog_3(-\ws) \Bigr)
\nonumber\\
&&\hskip-2.5cm\null
- 2 \, \ln|w|^2 \, \ln\frac{|1+w|^4}{|w|^2} 
           \biggl( \polylog_3\left(\frac{w}{1+w}\right)
                 + \polylog_3\left(\frac{\ws}{1+\ws}\right) \biggr)
\nonumber\\
&&\hskip-2.5cm\null
- 6 \, \ln|w|^2 \, \ln|1+w|^2 \, \ln\frac{|1+w|^2}{|w|^2}
  \Bigl( \polylog_2(-w) + \polylog_2(-\ws) \Bigr)
\nonumber\\
&&\hskip-2.5cm\null
+ \frac{5}{3} \, \ln^5|1+w|^2
- \frac{5}{2} \, \ln|w|^2 \, \ln^4|1+w|^2
+ \frac{4}{3} \, \ln^2|w|^2 \, \ln^3|1+w|^2
\nonumber\\
&&\hskip-2.5cm\null
- \ln|w|^2 \, \ln^2(1+w) \, \ln^2(1+\ws)
- 2 \, \ln^3|1+w|^2 \, \ln(1+w) \, \ln(1+\ws)
\nonumber\\
&&\hskip-2.5cm\null
+ \zeta_2 \, \ln|w|^2 \, \ln|1+w|^2 
     ( \ln|w|^2 - 3 \, \ln|1+w|^2 ) 
+ 4 \, \zeta_3 \, \ln|w|^2 \, \ln|1+w|^2 
- 48 \,\zeta_5 \Biggr\}
\nonumber\\
&&\hskip-2.5cm\null
+ \zeta_{3} \, d_1 \, g^{(2)}_1(w,\ws) 
+ \zeta_{2} \,  \gamma'' \, g^{(2)}_0(w,\ws) 
+ \zeta_2 \, d_2 \, \ln|1+w|^2 \, 
\ln\frac{|1+w|^2}{|w|^2} \, \ln\frac{|1+w|^4}{|w|^2} \,.
\label{g0wto0}
\eea
Note that although $\polylog_m(1/(1+w))$ has logarithmic branch-cut
behaviour near $w=0$, the combination
\be
\polylog_m\left(\frac{1}{1+w}\right)
- \frac{(-1)^m}{(m-1)!} \ln w \ln^{m-1}(1+w)
\label{goodpolylogcomb}
\ee
is well-behaved.  This property can be verified inductively by
differentiating with respect to $w$ and using
\be 
\frac{d}{dw} \polylog_m\left(\frac{1}{1+w}\right)
= - \frac{1}{1+w} \, \polylog_{m-1}\left(\frac{1}{1+w}\right)  \,.
\ee
After using the combination~(\ref{goodpolylogcomb})
in $g_0^{(3)}$, there are no other bare $\ln w$ terms; they all
come along with a $\ln w^*$ to form $\ln|w|^2$.  {Note that for $m=3$
one can use an identity to eliminate $\polylog_3(1/(1+w))$ in favor of
$\polylog_m(-w)$ and $\polylog_3(w/(1+w))$, but there is no such
identity for $m>3$.}

As was the case for $g^{(3)}_1$, all possible constraints will be
satisfied by a function proportional to the two-loop remainder
function, multiplied by $\zeta_2$.
This accounts for the term proportional to $g^{(2)}_0(w,\ws)$.  In
addition, we can multiply the two-loop leading-log multi-Regge
coefficient $g^{(2)}_1$ by $\zeta_3$, to get something with the right
transcendental degree and satisfying the above constraints.
Presumably its coefficient, $d_1$, can be fixed by additional
beyond-the-symbol information.  Finally, there is another degree-three
function satisfying all the constraints we imposed, with a coefficient
$d_2$ which we expect to be fixed in a similar fashion.  This
purely-logarithmic degree-three function is a linear combination of
the next-to-leading-log two-loop function $g^{(2)}_0$ and the
leading-log three-loop function $g^{(3)}_2$, as in~\eqn{g2wto0}.

The real part at next-to-next-to-leading-log level is given by
the degree-four function,
\bea
h^{(3)}_0(w,\ws) &=& \frac{1}{16} \Biggl\{
- \Bigl( 3 \, \ln |1+w|^2 - 2 \, \ln |w|^2 \Bigr) 
 \Bigl[ \polylog_3(-w) + \polylog_3(-\ws) \Bigr] 
\nonumber\\
&&\hskip-2.5cm\null
+ \ln |w|^2 \, \biggl[ \polylog_3\left(\frac{w}{1+w}\right)
      + \polylog_3\left(\frac{\ws}{1+\ws}\right) \biggr]
+ \frac{1}{2} \,  \ln |w|^2 \, \ln\frac{|1+w|^4}{|w|^2} 
 \Bigl[ \polylog_2(-w) + \polylog_2(-\ws) \Bigr]
\nonumber\\
&&\hskip-2.5cm\null
- \frac{1}{2} \, \ln^4 |1+w|^2
+ \frac{5}{6} \, \ln^3 |1+w|^2 \, \ln |w|^2
- \frac{1}{4} \, \ln^2 |1+w|^2 \, \ln^2 |w|^2
\nonumber\\
&&\hskip-2.5cm\null
+ \frac{1}{2} \, \ln |w|^2 \, \ln\frac{|1+w|^2}{|w|^2}
              \, \ln(1+w) \, \ln(1+\ws)
- 2 \, \zeta_3 \, \ln |1+w|^2 \Biggr\}
+ \zeta_2 \, \gamma''' \, g_1^{(2)} \,.
\label{h0wto0}
\eea
As was the case for \eqn{g1wto0}, the term containing an explicit 
$\zeta_3$ in \eqn{h0wto0} is fixed using the symmetries and the vanishing
of $h^{(3)}_0$ as $|w|\to0$.  There is an arbitrary constant
$\gamma'''$ multiplying $g_1^{(2)}$, but we will see shortly how to fix it.

In ref.~\cite{Lipatov:2010qf}, the scattering amplitude in the multi-Regge
limit was expressed as a sum of Regge pole and Mandelstam cut contributions. 
By using this representation, general formulae were obtained for the
multi-Regge limit of the remainder function in both $2\to4$ and $3\to3$
kinematics, in terms of a real function $f(\omega;x,y)$ characterizing
the partial waves entering the Mandelstam cut, 
\bea
\exp[ R_6 \, + \, i\pi\delta ] &=& \cos\pi\omega_{ab}
\, + \, i \int_{-i\infty}^{i\infty} \frac{d\omega}{2\pi i}
 \, f(\omega;x,y) \, e^{-i\pi\omega} \, |1-u|^{-\omega} \qquad (2\to4),
\label{gentwotofour}\\
\exp[ R_6 \, - \, i\pi\delta ] &=& \cos\pi\omega_{ab}
\, - \, i \int_{-i\infty}^{i\infty} \frac{d\omega}{2\pi i}
 \, f(\omega;x,y) \, |1-u|^{-\omega} \qquad \hskip0.99cm (3\to3).
\label{genthreetothree}
\eea
Here
\bea
\exp[ R_6 ] &=& 1 + a^2 \, R_6^{(2)} + a^3 \, R_6^{(3)} + \ldots,
\label{R6BLP}\\
\delta &=& - \frac{\gamma_K}{8} \, \ln \frac{|1+w|^4}{|w|^2} \,,
\label{deltaBLP}\\
\omega_{ab} &=& \frac{\gamma_K}{8} \, \ln |w|^2 \,,
\eea
and the cusp anomalous dimension $\gamma_K$ is given by 
\be
\gamma_K =  4 \, a - 4 \, \zeta_2 \, a^2 + 22 \, \zeta_4 \, a^3 + \ldots,
\label{gamma_cusp}
\ee
in terms of the coupling constant $a = g^2 N_c/(8\pi^2)$.
Note that the quantity appearing in
\eqns{gentwotofour}{genthreetothree} is the ratio of the full amplitude
(or Wilson loop) to the BDS ansatz, which according to our
conventions (see \eqns{WardIdSoln}{Rnnorm})
is the exponential of the remainder function.
The phase $\delta$ comes from the behavior of the BDS ansatz in the
multi-Regge limit, while $\omega_{ab}$ is derived from the Regge-pole
contribution.

Remarkably, the second term in \eqn{genthreetothree}, containing
$f(\omega;x,y)$, drops out when we take the real part of the equation,
leading to the all-loop-order relation for $3\to3$
kinematics~\cite{Bartels:2010tx},
\be
{\rm Re}\Bigl\{ \exp [ R_6 \, - \, i\pi\delta ] \Bigr\}
= \cos\pi\omega_{ab}
\qquad (3\to3)\,.
\label{allloopthreetothree}
\ee
The factor of $e^{-i\pi\omega}$ inside the integral in \eqn{gentwotofour}
prevents an analogously simple relation from holding for $2\to4$ scattering.

By using the results given above for the functions $g_r^{(l)}$ and $h_r^{(l)}$
through $l=3$, we can easily test \eqn{allloopthreetothree} at the
three-loop level.  We assemble the exponential of the remainder function, 
$\exp[R_6]$ in \eqn{R6BLP}, using 
\eqn{R6lregge} for $2\to4$ kinematics.
Then we apply the substitution~(\ref{lnomushift}),
followed by complex conjugation, to convert the result into the one for $3\to3$
kinematics.  Dressing the result with $e^{-i\pi\delta}$ and taking
the real part, we find that \eqn{allloopthreetothree} is satisfied precisely,
through three loops --- but only if we set $c=0$ in \eqn{g1wto0} for
$g^{(3)}_1$.
In addition we must fix the constant $\gamma'''$
in \eqn{h0wto0} for $h_0^{(3)}$ to the value,
\be
\gamma''' = \frac{9}{4} + \frac{\gamma'}{2} \,.
\label{determine_gamma_ppp}
\ee
The imaginary part $g^{(3)}_1$ for $2\to4$ kinematics contributes
to the real part of $R_6^{(3)}$ for $3\to3$ kinematics because of the
substitution~(\ref{lnomushift}) and the fact that $g^{(3)}_1$ is
multiplied by $\ln(1-u)$ in \eqn{R6lregge}.  In fact, the only function
that does not enter \eqn{allloopthreetothree} is the degree-five function
$g^{(3)}_0$, because it is from the imaginary part and has no $\ln(1-u)$
multiplying it.  Hence \eqn{allloopthreetothree} is a powerful check
on our results.

The $c=0$ constraint imposed by \eqn{allloopthreetothree} also arises
from considering restrictions on the final entry of the symbol, as we
shall do in the next section.


\section{Constraints on the final entry of the symbol}
\label{finalentry}

We have shown that within the specific ansatz~(\ref{ansatz}) we were
able to write the symbol of the three-loop remainder function in the
form
\be 
\mathcal{S}(R_6^{(3)}) = \mathcal{S}(X)
 + \sum_{i=1}^{24} \alpha_i \, \mathcal{S}(f_i)\,.
\ee
There are 24 unfixed parameters $\alpha_i$, after imposing all of the
constraints we have outlined,
including the constraints coming from the multi-Regge
limit~(\ref{reggeconstraints}).  Moreover, by examining the symbol
$\mathcal{S}(X)$ we find that it is possible to write it so that the
final entries are drawn from the following set,
\be
\Bigl\{ \frac{u}{1-u} , \frac{v}{1-v}, \frac{w}{1-w} , 
       y_u, y_v, y_w \Bigr\}\,.
\ee
The same restriction is true for the symbol of the full remainder function
$R_6^{(2)}$ at two loops, given in \eqn{R62symb}.
As mentioned above, it has been suggested~\cite{CaronHuot:2011ky}
that this fact is related to a supersymmetric formulation of
the Wilson loop; and similar restrictions appear~\cite{DDHtoappear}
in differential equations~\cite{Drummond:2010cz,Dixon:2011ng,DelDuca:2011wh}
for integrals related to scattering amplitudes~\cite{ArkaniHamed:2010kv}.
It is reasonable to think that the full symbol
$\mathcal{S}(R_6^{(3)})$ should obey this condition, including the
ambiguities $\mathcal{S}(f_i)$. In fact, it is possible to impose this
condition on the remaining ambiguities, leaving just two free
parameters,
\be
\mathcal{S}(R_6^{(3)}) = \mathcal{S}(X)
 + \alpha_1 \, \mathcal{S}(f_1) + \alpha_2 \, \mathcal{S}(f_2)\,.
\label{symbolR63}
\ee
The fact that this form for the symbol is consistent with the known
Regge behaviour is highly non-trivial.  Indeed, one can adopt
the constraint on the final entries from the beginning.  In this case,
after imposing the OPE constraints, the triple-log in the multi-Regge limit
vanishes automatically, and the leading-log contribution $g_2^{(3)}$ is
uniquely fixed to agree with the prediction of
refs.~\cite{Bartels:2010tx,Bartels:2011xy}.  Finally, the single
remaining free parameter in the multi-Regge limit (which appears in
the function $g_1^{(3)}$ in~\eqn{g1wto0}) is fixed,
\be
c=0\,,
\ee
leaving a completely unambiguous prediction for the symbol of
$g^{(3)}_1$ in this limit (the symbol for $g^{(3)}_0$ was already
fixed unambiguously).  It is reassuring that the same vanishing value
for $c$ is also dictated by the relation~(\ref{allloopthreetothree})
for the multi-Regge limit for $3\to3$ kinematics.

The symbol ${\mathcal S}(f_1)$ is extremely simple:  It is entirely composed
from the entries $\{u,v,w,1-u,1-v,1-w\}$; the
square-root containing $y$ variables in \eqn{ansatz} do not
appear in ${\mathcal S}(f_1)$.  This property makes it
straightforward to find an explicit function $f_1$, which has
the symbol $\mathcal{S}(f_1)$.  The function can be written in the form,
\be
f_1(u,v,w) = h(u)h(v) + h(u)h(w) + h(v)h(w) + k(u) + k(v) + k(w)\,.
\ee
Here the single-variable functions $h$ and $k$ are given by
\begin{align}
h(u)\ = &\ \tfrac{1}{3} \, \log^3 u 
+ \log u \, {\rm Li}_2(1-u) - {\rm Li}_3(1-u) - 2 \, {\rm Li}_3(1- 1/u) \,, \\
k(u)\ = & - \log^3 u \, H_{3}
+ \tfrac{3}{2} \, \log^2 u \, ( H_{4} - H_{2, 2}  - 4 \, H_{3, 1} ) \notag \\
& - \log u \, ( H_{2, 3}  - 6 \, H_{4, 1} + H_{2, 1, 2}
              + 6 \, H_{2, 2, 1} + 18 \, H_{3, 1, 1} ) \notag \\
& + 3 \, H_{2, 4} + 4 \, H_{3, 3} + 3 \, H_{4, 2}
+ H_{2, 1, 3} - H_{2, 2, 2} - 2 \, H_{2, 3, 1}  \notag \\
& - 2 \, H_{3, 1, 2} + 9 \, H_{4, 1, 1} - 2 \, H_{2, 1, 2, 1}
  - 9 \, H_{2, 2, 1, 1} - 24 \, H_{3, 1, 1, 1}
 \,.
\end{align}
The arguments of the harmonic polylogarithms appearing in $k(u)$ are all
$(1-u)$ and have been suppressed to save space.  A subscript $m$ stands for
$m-1$ zero entries followed by a single 1 entry~\cite{Remiddi:1999ew};
so for example $H_{3, 1, 2} = H_{0,0,1,1,0,1}(1-u)$.

The function $f_1$ above has been chosen so that it obeys
\be
\partial_u f_1 = \frac{1}{u(1-u)} (\text{pure function})\,.
\label{uderiv}
\ee
The fact that taking the derivative yields a pure function with the
particular $1/(u(1-u))$ prefactor is the functional consequence of the
final-entry condition on the symbol.  The function $f_1$ is
real-valued in the Euclidean region but does not vanish in the
collinear limit.  It only vanishes up to terms involving explicit
appearances of $\zeta_2$ ($\pi^2$) and $\zeta_3$, which is what is
guaranteed by the form of its symbol.  In fact, already at the
$\zeta_2$ level we find that $f_1$ is divergent in this limit,
\begin{align}
\lim_{w \rightarrow 0} f_1 = 
\zeta_2 \Bigl[ &\log w \Bigl( \tfrac{1}{2} \log u \log^2(1-u) 
+ \log u {\rm Li}_2(u) +  2 \log(1-u) {\rm Li}_2(u) - 3 {\rm Li}_3(u)
 + 3 H_{2,1}(u) \Bigr) \notag \\
 & +  \text{finite}\Bigr] + \zeta_3 \Bigl[ \ldots \Bigr] \,.
\end{align}
In fact there exists no degree 4 function with a symbol within our
ansatz, and also obeying the property~(\ref{uderiv}), which could be used
to remove this divergence in the collinear limit.  This fact suggests that
if we insist on preserving the functional consequence of the final
entry condition~(\ref{uderiv}), beyond the level of the symbol,
then there will be additional constraints on the parameter $\alpha_1$
when completing the symbol $\mathcal{S}(R_6^{(3)})$ to a genuine function.

The function $f_2$ is intermediate in complexity between $f_1$ and $X$.
Its symbol contains terms with up to two $y$-variable entries, while
the symbol for $X$ has terms with four $y$-variable entries.  Files
containing the symbols for $f_1$, $f_2$ and $X$, as well as the
symbols of the functions characterizing the multi-Regge
limit (which for $c=0$ come entirely from $X$), are provided as auxiliary
material for this paper.

We leave to later work an explicit construction of functions associated
with the other symbols, particularly ${\mathcal S}(X)$ and
${\mathcal S}(f_2)$.  However, we can already say some things about 
the full three-loop remainder function.
In particular, for any values of $\alpha_1$ and $\alpha_2$,
it is impossible to represent its symbol by a function
given in terms of (products of) only single-variable harmonic
polylogarithms $H_{\vec{w}}(x)$, whose weight vectors $\vec{w}$
contain only the entries $0$ and $1$.  As a corollary, it
is not possible to represent the symbol by a function given purely in
terms of the classical polylog functions ${\rm Li}_n(x)$, for any
choices of $x$.  This result can be obtained by performing symmetry
operations similar to those described in ref.~\cite{Goncharov:2010jf}.
It is sufficient, and a bit simpler, to test not the full symbol, but
a particular piece of it.  We take the double discontinuity in $w$,
and then set $w\to0$, using the relations~(\ref{y_as_wto0}).
This symbol is given by
\bea
 && {\mathcal S}(\Delta_w \Delta_w X)|_{w\to0}
\ =\ \frac{1}{8} \Biggl\{
  u \otimes u \otimes \biggl[ - (1-u) \otimes \frac{uv}{(1-u)(1-u-v)}
                            + v \otimes \frac{1-v}{1-u-v} \biggr]
\nonumber \\ &&\hskip0.0cm \null
+ u \otimes (1-u) \otimes \biggl[
   \frac{1-u}{(1-u-v)^2} \otimes \frac{uv}{(1-u)(1-u-v)}
   + \frac{u}{(1-u-v)^2} \otimes \frac{(1-u)^2(1-v)}{(1-u-v)^3}
\nonumber \\ &&\hskip3.1cm \null
   + v(1-u-v) \otimes \frac{(1-u)(1-v)}{(1-u-v)^2} \biggr]
\nonumber\\
&&\null
+ u \otimes v \otimes \biggl[
  - 2\ \frac{(1-u)(1-v)}{1-u-v} \otimes \frac{uv}{(1-u-v)^2}
  + u(1-u-v) \otimes \frac{1-v}{1-u-v} 
\nonumber \\ &&\hskip2.1cm \null
  + v(1-u-v) \otimes \frac{1-u}{1-u-v} \biggr]
\Biggr\}\ +\ (u \longleftrightarrow v) \,.
\label{Delta_w_Delta_w_X_weq0}
\eea
We replace each term of the form $a \otimes b \otimes c \otimes d$
in this expression with the following
antisymmetrisation~\cite{Goncharov:2010jf}:
\be \Bigl[
(a \otimes b \otimes c \otimes d 
- (c \leftrightarrow d)) - (a \leftrightarrow b) \Bigr]
- \Bigl[(a,b) \leftrightarrow (c,d) \Bigr]\,.
\label{degreefourLitest}
\ee
We find that \eqn{Delta_w_Delta_w_X_weq0} is nonvanishing under
this operation. The symbol of a degree four function constructed solely
from products of single-variable harmonic polylogarithms with labels
0 and 1 (which includes all ${\rm Li}_n$ functions) vanishes under this
operation.  Hence $(\Delta_w \Delta_w X)_{w\to0}$, and also $X$ itself,
must include functions beyond this class.

We have also performed a similar test on the full degree
six function.  Given a degree six symbol
which is a sum of terms of the form
$a \otimes b \otimes c \otimes d \otimes e \otimes f$, we replace each
term with the following antisymmetrisation,
\be \Bigl[\bigl( 
(a \otimes b \otimes c \otimes d \otimes e \otimes f 
- (e \leftrightarrow f)) - (c \leftrightarrow d) \bigr)
 - (a \leftrightarrow b)\Bigr]
- \Bigl[(a,b) \leftrightarrow (e,f) \Bigr]\,.
\ee
The symbol of a degree six function constructed solely from products of
single-variable harmonic polylogarithms with labels 0 and 1
vanishes under this operation.  We find that ${\mathcal S}(X)$
does not vanish under this operation, so again we conclude that
$X$ must include functions beyond this class.


\section{Conclusions and outlook}
\label{sect-conclusions}

In this paper we determined the symbol of the remainder function for
the three-loop hexagon Wilson loop, or six-point MHV scattering
amplitude, in planar $\cN=4$ super-Yang-Mills theory, up to a few
undetermined constants.  There are 26 such constants in a more general
ansatz, but this number drops to just two if a final-entry restriction
is imposed on the symbol.  The OPE expansion, as analysed in
refs.~\cite{Alday:2010ku,Gaiotto:2010fk,Gaiotto:2011dt},
provides a powerful constraint for this problem, which is straightforward
to implement with the aid of symbols.  In particular, we uniquely determined
the symbol $\mathcal{S}(X)$ for the part of the three-loop remainder
function that has a leading discontinuity.

In the multi-Regge limit,
all but one of the symbol-level constants drop out (all of them
drop out when we impose the final-entry restriction).  In this limit,
we are able to complete the symbols for the coefficients in the
logarithmic expansion into full analytic functions of degree
3, 4 and 5.  These functions depend on two variables, yet they can
all be expressed in terms of classical polylogarithms.  Three of these
functions represent new predictions for the behaviour of the amplitude
in the multi-Regge limit.  We found confirmation of the final-entry 
restriction by testing an all-order relation for the remainder function
in multi-Regge kinematics for $3\to3$ scattering.

Although only classical polylogarithms appear in the multi-Regge limit,
we could use our symbol to show that for more
generic kinematics, the three-loop remainder function cannot be
expressed solely in terms of classical polylogarithms.  Clearly it
is an important task to complete the terms in this symbol into full
functions.  For $f_1$, one of the two terms that we could not fix using the
leading discontinuity (assuming the final-entry restriction),
we were able to accomplish this task.  This function is particularly
simple due to the fact that its symbol does not depend on the $y$ variables,
but only on $\{u,v,w,1-u,1-v,1-w\}$.  It factorises into single-variable
functions constructed out of harmonic polylogarithms.

The next simplest component is $f_2$.  It is only quadratic in the $y$
variables, so in some sense it is not much more complicated than the
two-loop remainder function, although it is of degree six instead of four.
The most complicated term is $X$, which is quartic in the $y$ variables.
However, we are optimistic that a relatively compact representation for it,
as well as $f_2$, may be possible to find.  We are also encouraged by
the fact that the collinear limits of $f_1$, which diverge beyond the
symbol level, cannot be repaired within functions obeying the final-entry
restriction.  This fact suggests that the repair may come instead through 
the collinear behaviour of $X$ and $f_2$, which could in turn fix one or both
of the arbitrary constants $\alpha_1$ and $\alpha_2$.
It would be remarkable if the three-loop remainder function could be
completely determined, or perhaps up to a single ambiguity associated with the
two-loop remainder function, without ever directly evaluating a single
loop integral, for either a Wilson loop or a scattering amplitude.


\section*{Acknowledgements}
We would like to thank Harald Dorn and Thomas Gehrmann for useful
discussions. J.M.H. is grateful to the KITP, Santa Barbara, for hospitality
while this work was carried out.  This work was supported in part by
the National Science Foundation under Grant No. PHY05-51164,
and by the US Department of Energy under contract DE--AC02--76SF00515.


\section*{Appendix A}
\label{KinRelationAppendix}

In this appendix we provide handy equations for relating various
differential operators in term of the variables $\tau$, $\sigma$ and $\phi$
to those in terms of the cross ratio variables $u$, $v$ and $w$.
From \eqn{uvwtausigmaphi} we have the auxiliary relations
\bea
\frac{1-u-v-w}{1-v} &=& \frac{\cos\phi}{\cosh\sigma \cosh\tau + \cos\phi} \,,
\qquad
\frac{4uvw}{(1-v)^2} = \frac{1}{(\cosh\sigma \cosh\tau + \cos\phi)^2} \,,
\label{uvwrels1}\\
\frac{\sqrt{\Delta}}{1-v} &=&
\frac{i \, \sin\phi}{\cosh\sigma \cosh\tau + \cos\phi} \,,
\qquad
\tanh\tau = \sqrt{1-v} \,.
\label{uvwrels2}
\eea
Using these relations, it is simple to show that
\bea
&&\frac{1}{u} \frac{\partial u}{\partial\tau}
 = \frac{1}{w} \frac{\partial w}{\partial\tau} 
 = \frac{1-u+v-w}{\sqrt{1-v}} \,, \qquad
\frac{1}{v}\frac{\partial v}{\partial\tau} = -2\sqrt{1-v} \,,
\label{difftau}\\
&&\frac{1}{u} \frac{\partial u}{\partial\sigma}
 = \frac{1-u-v+w}{1-v} \,, \qquad
\frac{1}{w} \frac{\partial w}{\partial\sigma} 
 = - \frac{1+u-v-w}{1-v} \,, \qquad
\frac{\partial v}{\partial\sigma} = 0 \,,
\label{diffsigma}\\
&&\frac{1}{u} \frac{\partial u}{\partial\phi}
 = \frac{1}{w} \frac{\partial w}{\partial\phi} 
 = \frac{1}{i} \frac{\sqrt{\Delta}}{1-v} \,,
\qquad
\frac{\partial v}{\partial\phi} = 0 \,.
\label{diffphi}
\eea

Then the operators differentiating with respect to $\tau$,
$\sigma$ and $\phi$ are
\bea
\partial_\tau &=& \frac{1}{\sqrt{1-v}}
\Bigl[ -2(1-v)v \partial_v
 + (1-u+v-w) ( u\partial_u + w\partial_w ) \Bigr] \,,
\label{diffwrttau}\\
\partial_\sigma &=& \frac{1}{1-v}
\Bigl[ (1-u-v+w) u\partial_u - (1+u-v-w) w\partial_w \Bigr] \,,
\label{diffwrtsigma}\\
\partial_\phi &=& \frac{\sqrt{\Delta}}{i(1-v)} 
  ( u\partial_u + w\partial_w ) \,.
\label{diffwrtphi}
\eea
Inserting these expressions into the form for $\mathcal{D}_\pm$ given
in ref.~\cite{Gaiotto:2011dt},
\be
\mathcal{D}_\pm = \partial_\tau^2 + 2 \, \coth(2\tau) \, \partial_\tau
   + {\rm sech}^2\tau \, \partial_\sigma^2
   + \partial_\phi ( \partial_\phi \mp 2i ) \,,
\label{DpmGMSV}
\ee
it is straightforward to obtain the form in terms of $u$, $v$ and $w$
given in \eqn{Dpmuvw}.  Similarly, the operator
$\Box = -(\partial_\sigma^2 + \partial_\phi^2)$ is found to have
the form given in \eqn{boxop}.


\section*{Appendix B}
\label{yuvwAppendix}

Although the $y$ variables are constructed using square roots
of the original cross ratios $u$, $v$ and $w$, the cross ratios
themselves are rational combinations of the variables $y_u$, $y_v$
and $y_w$.  The explicit relations are,
\bea
&& u = \frac{y_u(1-y_v)(1-y_w)}{(1-y_w y_u)(1-y_u y_v)} \,, \quad
v = \frac{y_v (1-y_w) (1-y_u)}{(1-y_u y_v) (1-y_v y_w)} \,, \quad
w = \frac{y_w (1-y_u) (1-y_v)}{(1-y_v y_w) (1-y_w y_u)} \,,~~ \\
&& 1-u = \frac{(1-y_u) (1-y_u y_v y_w)}{(1-y_w y_u) (1-y_u y_v)} \,, \qquad
   1-v = \frac{(1-y_v) (1-y_u y_v y_w)}{(1-y_u y_v) (1-y_v y_w)} \,, \\
&& 1-w = \frac{(1-y_w) (1-y_u y_v y_w)}{(1-y_v y_w) (1-y_w y_u)} \,, \qquad
\sqrt{\Delta} = \frac{(1-y_u) (1-y_v) (1-y_w) (1-y_u y_v y_w)}
                       {(1-y_u y_v) (1-y_v y_w) (1-y_w y_u)} \,,
\eea
where we have picked a particular branch of $\sqrt{\Delta}$.
These formulas are also useful in the multi-Regge limit.
The limit~(\ref{multireggekinematics}) corresponds to taking
$y_u\to1$, $y_v\to\ty_v$, $y_w\to\ty_w$.  We find in the limit,
\be
x = \frac{\ty_v (1-\ty_w)^2}{(1-\ty_v \ty_w)^2} \,, \qquad
y = \frac{\ty_w (1-\ty_v)^2}{(1-\ty_v \ty_w)^2} \,, \qquad
\sqrt{\tilde\Delta} = \frac{(1-\ty_v)(1-\ty_w)}{1-\ty_v\ty_w} \,.
\ee
The variables $w$ and $\ws$ used in the multi-Regge limit,
defined in \eqn{wwsdef}, are also rational combinations of
$\ty_v$ and $\ty_w$:
\be
w = \frac{1-\ty_v}{\ty_v(1-\ty_w)} \,, \qquad
\ws = \frac{\ty_w(1-\ty_v)}{1-\ty_w} \,.
\ee
Inverting these equations gives the expressions for $\ty_v$ and $\ty_w$
in terms of $w$ and $\ws$ given in \eqns{yvtilde}{ywtilde}.


\section*{Appendix C}
\label{gsymformAppendix}

Here we write the three-loop functions $g^{(3)}_r$ and $h^{(3)}_r$
in a form that makes the $w$ inversion
and $w \leftrightarrow \ws$ symmetries manifest.  To do so, we
introduce functions $\hat{g}^{(l)}_r(w,\ws)$ and $\hat{h}^{(l)}_r(w,\ws)$ 
such that the sum over
images under the two symmetries yields the full functions:
\bea
g^{(l)}_r(w,\ws) &=& \hat{g}^{(l)}_r(w,\ws) + \hat{g}^{(l)}_r(\ws,w)
         + \hat{g}^{(l)}_r(1/w,1/\ws) + \hat{g}^{(l)}_r(1/\ws,1/w) \,,
\label{gsymdecomp}\\
h^{(l)}_r(w,\ws) &=& \hat{h}^{(l)}_r(w,\ws) + \hat{h}^{(l)}_r(\ws,w)
         + \hat{h}^{(l)}_r(1/w,1/\ws) + \hat{h}^{(l)}_r(1/\ws,1/w) \,.
\label{hsymdecomp}
\eea
We find that
\bea
\hat{g}^{(3)}_{2}(w,\ws) &=& 
- \frac{1}{32} \biggl[ 2 \, \polylog_{3}(-w)
     - \ln|w|^2 \, \polylog_{2}(-w) \nonumber\\
&&\hskip1cm\null - \frac{1}{12} \ln^2|1+w|^2
 \left( \ln|1+w|^2 - 9 \, \ln\frac{|1+w|^2}{|w|^2} \right) \biggr]
\,,
\eea
which agrees with \eqn{g2wto0} and with ref.~\cite{Lipatov:2010ad}
after the use of a few polylogarithm identities.
Similarly, $h^{(3)}_1$ can be written symmetrically using
\bea
\hat{h}^{(3)}_{1}(w,\ws) &=& 
- \frac{1}{32} \biggl[ 2 \, \polylog_{3}(-w)
     - \ln|w|^2 \, \polylog_{2}(-w) \nonumber\\
&&\hskip1cm\null - \frac{1}{12} \ln^2|1+w|^2
 \left( \ln|1+w|^2 + 3 \, \ln\frac{|1+w|^2}{|w|^2} \right) \biggr]
\,.
\eea

The new functions found in this paper are $g^{(3)}_1$, $g^{(3)}_0$
and $h^{(3)}_0$.  For $g^{(3)}_1$ the symmetric form uses
\bea \label{g1}
\hat{g}^{(3)}_{1}(w,\ws) &=& 
- \frac{1}{32} \biggl\{
\nonumber\\
&&\hskip-2.2cm\null
4 \Bigl( 5 \, \ln|1+w|^2 - 2 \, \ln|w|^2 \Bigl) 
  \, \polylog_{3}\left(\frac{1}{1+w}\right)
+ 3 \, \ln|w|^2 \, \ln\frac{|1+w|^4}{|w|^2} \, \polylog_{2}(-w)
\nonumber\\
&&\hskip-2.5cm\null
+ \frac{3}{16} \biggl[ \ln^4|w|^2
 + 8 \, \ln^2|w|^2 \, \ln(1+\ws) \, \ln \frac{1+w}{w}
  + 2 \, \ln|w|^2 \, \ln\frac{w}{\ws} \, \ln^2\frac{(1+w)^2}{w} \biggr]
\nonumber\\
&&\hskip-2.5cm\null
- 5 \, \ln\frac{|1+w|^2}{|w|^2} \, \ln|1+w|^2 
    \, \ln(1+w) \, \ln\frac{1+w}{w}
+ \frac{3}{2} \, \zeta_{2} \, \ln^2|w|^2
- 8 \, \zeta_{3} \, \ln|1+w|^2 \biggr\}
\nonumber\\
&&\hskip-2.5cm\null
+ \frac{\zeta_{2}}{4} \, \gamma' \, g^{(2)}_1(w,\ws) \, 
   + \, c \, \hat g_1^a \,,
\eea
The constant $c$ multiplies the function,
\bea
\hat g_1^a &=& - 4 \, \ln|1+w|^2 \, \polylog_{3}\left(\frac{w}{1+w}\right)
+ \polylog_2(-w) \Bigl[ \polylog_2(-w) - \polylog_2(-\ws)
                  + \ln |w|^2 \, \ln \frac{1+w}{1+\ws} \Bigr]
\nonumber\\
&&\hskip-1.0cm\null
+ \frac{1}{24} \, \ln^4|1+w|^2
- \frac{1}{4} \, \ln |w|^2 \, \ln^2|1+w|^2 \, \ln\frac{|1+w|^2}{|w|^2}
\nonumber\\
&&\hskip-1.0cm\null
+ \frac{1}{2} \, \ln |w|^2 \, \ln w \, \ln(1+w) \, \ln\frac{1+w}{\ws}
+ \frac{1}{3} \,  \ln|1+w|^2 \, \ln^2(1+w) \, ( 2 \, \ln(1+w) - 3 \, \ln w )
\nonumber\\
&&\hskip-1.0cm\null
- \zeta_2 \,  \ln|1+w|^2 \, \ln\frac{|1+w|^2}{|w|^2}
+ 4 \, \zeta_3 \, \ln|1+w|^2 \,.
\eea
Recall that $c=0$ if we impose either the all-loop-order
prediction for $3\to3$ scattering~\cite{Bartels:2010tx}, or
an additional constraint on the form of the final entries in the
symbol of $R_6^{(3)}$, as described in section~\ref{finalentry}.

The function entering the symmetric form for $g^{(3)}_0$ is
\bea
\hat{g}^{(3)}_{0}(w,\ws) &=& 
- \frac{1}{32} \biggl\{
\nonumber\\
&&\hskip-2.5cm\null
- 30 \, \Bigl( 2 \, \polylog_{5}(-w) - \ln|w|^2 \, \polylog_{4}(-w) \Bigr)
+ 12 \, \biggl( 2 \, \polylog_{5}\left(\frac{1}{1+w}\right)
              + \ln|1+w|^2 \, \polylog_{4}\left(\frac{1}{1+w}\right) \biggr)
\nonumber\\
&&\hskip-2.5cm\null
+ \Bigl( - 6 \, \ln^2|w|^2 - 4 \, \ln|w|^2 \, \ln|1+w|^2
  + 5 \, \ln^2|1+w|^2 \Bigr) \, \polylog_{3}(-w)
\nonumber\\
&&\hskip-2.5cm\null
+ \ln|w|^2 \, \ln\frac{|1+w|^4}{|w|^2} 
  \, \polylog_{3}\left(\frac{1}{1+w}\right)
- 3 \, \ln|w|^2 \, \ln|1+w|^2 \, \ln\frac{|1+w|^2}{|w|^2} \, \polylog_{2}(-w)
\nonumber\\
&&\hskip-2.5cm\null
- \frac{1}{48} \, \ln\frac{|1+w|^4}{|w|^2}
      \, \biggl( \ln^4|1+w|^2 + \ln^4 \frac{|1+w|^2}{|w|^2}
        - 9 \, \ln^2|1+w|^2 \, \ln^2 \frac{|1+w|^2}{|w|^2} \biggr)
\nonumber\\
&&\hskip-2.5cm\null
+ \frac{1}{32} \, \ln^2 \frac{w}{\ws} \, \ln|w|^2 \, \ln^2|1+w|^2
\nonumber\\
&&\hskip-2.5cm\null
- \frac{1}{32} \, \ln\frac{1+w}{1+\ws}
        \, \biggl( \ln^2|w|^2 + 2 \, \ln|1+w|^2 
    \, \ln\frac{|1+w|^2}{|w|^2} \biggr)
\nonumber\\
&&\hskip-1.5cm\null
   \times \biggl( 2 \, \ln\frac{w}{\ws} \, \ln|1+w|^2
          - \ln\frac{1+w}{1+\ws} \, \ln\frac{|1+w|^4}{|w|^2} \biggr) 
+ \frac{1}{2} \, \zeta_{3} \, \ln^2|w|^2
- 12 \, \zeta_{5} \biggr\}
\nonumber\\
&&\hskip-2.5cm\null
+ \frac{\zeta_{3}}{4} \, d_1 \, g^{(2)}_1(w,\ws) 
+ \frac{\zeta_{2}}{4} \,  \gamma'' \, g^{(2)}_0(w,\ws) 
+ \frac{\zeta_2}{4} \, d_2 \, \ln|1+w|^2 
  \, \ln\frac{|1+w|^2}{|w|^2} \, \ln\frac{|1+w|^4}{|w|^2} \,.
\eea

Finally, the function needed to write $h^{(3)}_0$ in a symmetric form is
\bea \label{h1sym}
\hat{h}^{(3)}_{0}(w,\ws) &=& 
\frac{1}{128} \biggl\{
\nonumber\\
&&\hskip-2.2cm\null
8 \Bigl( 3 \, \ln|1+w|^2 - 2 \, \ln|w|^2 \Bigl) 
  \, \polylog_{3}\left(\frac{1}{1+w}\right)
+ 2 \, \ln|w|^2 \, \ln\frac{|1+w|^4}{|w|^2} \, \polylog_{2}(-w)
\nonumber\\
&&\hskip-2.5cm\null
+ \frac{1}{8} \biggl[ \ln^4|w|^2
 + 8 \, \ln^2|w|^2 \, \ln(1+\ws) \, \ln \frac{1+w}{w}
  + 2 \, \ln|w|^2 \, \ln\frac{w}{\ws} \, \ln^2\frac{(1+w)^2}{w} \biggr]
\nonumber\\
&&\hskip-2.5cm\null
- 6 \, \ln\frac{|1+w|^2}{|w|^2} \, \ln|1+w|^2 
    \, \ln(1+w) \, \ln\frac{1+w}{w}
+ \zeta_{2} \, \ln^2|w|^2
- 16 \, \zeta_{3} \, \ln|1+w|^2 \biggr\}
\nonumber\\
&&\hskip-2.5cm\null
+ \frac{\zeta_{2}}{4} \, \biggl( \frac{3}{4} + \gamma''' \biggr)
  \, g^{(2)}_1(w,\ws) \,.
\eea
Polylogarithm identities are required to go between the manifestly
symmetric forms of the functions $g_r^{(3)}(w,w^*)$ and
$h_r^{(3)}(w,w^*)$ given in this appendix,
and the forms~(\ref{g2wto0}), (\ref{g1wto0}), (\ref{g1awto0}),
(\ref{h1}), (\ref{g0wto0}) and (\ref{h0wto0}) given in the main text,
which have good behaviour as $w\to0$.


\bibliographystyle{nb}

\end{document}